\documentclass{elsart}

\usepackage{graphics}
\usepackage{graphicx}
\usepackage{lineno}
\usepackage{amssymb,amsmath}

\begin{document}

\begin{frontmatter}



  \title{Position resolution and particle identification with the ATLAS EM
  calorimeter} 


%\author{ATLAS Electromagnetic Liquid Argon Calorimeter Group} %\\[1cm]
%\address{}

\author[Annecy]{J.~Colas},
\author[Annecy]{L.~Di Ciaccio},
\author[Annecy]{M.~El~Kacimi\thanksref{ElKacimi}},
\author[Annecy]{O.~Gaumer},
\author[Annecy]{M.~Gouan\`ere},
\author[Annecy]{D.~Goujdami\thanksref{ElKacimi}},
\author[Annecy]{R.~Lafaye},
\author[Annecy]{C.~Le Maner},
\author[Annecy]{L.~Neukermans},
\author[Annecy]{P.~Perrodo},
% ...
%\author[StonyBrook]{J.~Egdemir},
%\author[StonyBrook]{R.~Engelmann},
%\author[StonyBrook]{J.~Hoffman},
%\author[StonyBrook]{R.~McCarthy},
%\author[StonyBrook]{M.~Rijssenbeek},
%\author[StonyBrook]{J. Steffens},

\address[Annecy]{Laboratoire de Physique de Particules (LAPP), 
IN2P3-CNRS, F-74941~Annecy-le-Vieux~Cedex, France.} 
\address[Brookhaven]{Brookhaven National Laboratory (BNL), Upton,
  NY~11973-5000, USA.}
\address[Casablanca]{Facult\'{e} des Sciences A\"{\i}n Chock, Casablanca,
  Morocco.} 
\address[Dallas]{Southern Methodist University, Dallas, Texas 75275-0175,
  USA.} 
\address[CERN]{European Laboratory for Particle Physics (CERN),
  CH-1211~Geneva~23, Switzerland.} 
%\address[Geneve]{Universit\'e de Gen\`eve, CH-1211~Geneva~4, 
%Switzerland.} 
\address[Grenoble]{Laboratoire de Physique Subatomique et de Cosmologie,
  Universit\'e Joseph Fourier, IN2P3-CNRS, F-38026~Grenoble, France.}
\address[Nevis]{Nevis Laboratories, Columbia University, Irvington, 
  NY~10533, USA.} 
\address[Madrid]{Physics Department, Universidad Aut\'{o}noma de Madrid,
  Spain.} 
\address[Marseille]{Centre de Physique des Particules de Marseille,
  Univ. M\'{e}diterran\'{e}e, IN2P3-CNRS, F-13288~Marseille, France.}
\address[Milano]{Dipartimento di Fisica dell'Universit\`{a} di Milano and 
  INFN, I-20133~Milano, Italy.} 
\address[Novosibirsk]{Budker Institute of Nuclear Physics,
  RU-630090~Novosibirsk, Russia.} 
\address[Orsay]{Laboratoire de l'Acc\'{e}l\'{e}rateur Lin\'{e}aire,
  Universit\'{e} de Paris-Sud, IN2P3-CNRS, F-91898~Orsay~Cedex, France.}
\address[Oujda]{Laboratoire de Physique Theorique et de Physique des
  Particules, Universit\'e Mohammed Premier, Oujda, Morocco.}
\address[Jussieu]{Universit\'es Paris VI et VII, Laboratoire de Physique
  Nucl\'eaire et de Hautes Energies, F-75252 Paris, France.} 
%\address[Pittsburgh]{Department of Physics and Astronomy, University of
%  Pittsburgh, Pittsburgh, PA~15260, USA.} 
\address[Rabat1]{Facult\'e des Sciences and 
  Centre National de l'\'Energie des Sciences et des Techniques
  Nucl\'eaires, Rabat, Morocco.}
\address[Rabat]{Universit\'e Mohamed V, Facult\'e des Sciences, Rabat, Morocco.}
\address[Saclay]{CEA, DAPNIA/Service de Physique des Particules, 
  CE-Saclay, F-91191~Gif-sur-Yvette~Cedex, France.}
\address[Stockholm]{Royal Institute of Technology, Stockholm, Sweden.}
%\address[StonyBrook]{State University of New York, Stony Brook,  New York
%  11794, USA.} 

\thanks[ElKacimi]{Visitor from LPHEA, FSSM-Marrakech (Morroco).}
\thanks[Barberio]{Now at university of Melbourne, Australia.}
\thanks[Beck]{Now at Niels Bohr Institute, Copenhagen.}
\thanks[Deceased]{Deceased.}
\thanks[Toth]{Also at KFKI, Budapest, Hungary.
  Supported by the MAE, France, the
  HNCfTD (Contract F15-00) and the Hungarian OTKA (Contract T037350).}
\thanks[cauthor]{E-mail: roberto.sacco@cern.ch.}
\thanks[Sacco]{Now at Queen Mary, University of London.} 
%Mile End Road, London
%E1 4NS.}

%\thanks[Garcia]{Now at "Instituto Nicolas Cabrera", U.A.M. Madrid.}
%\thanks[Alexa]{Also at Institute of Atomics Physics, National Institute
%  for Physics and Nuclear Engineering IFIN-HH, Bucharest, Romania.}

\begin{abstract}
%The results of some analyses relying on the fine lateral segmentation 
%of the  
%ATLAS EM calorimeter are presented.
In the years between 2000 and 2002 several pre-series and series 
modules of the ATLAS EM barrel and end-cap calorimeter were exposed to 
electron, photon
and pion beams. The performance of the calorimeter with respect to its 
finely segmented first sampling has been studied. The
polar angle resolution has been found to be in the range
$50-60~(\mathrm{mrad})/\sqrt{E~(\mathrm{GeV})}$. 
The $\pi^0$ rejection has
been measured to be about 3.5 for 90\% photon selection
efficiency at $p_T=50$~GeV/c. e-$\pi$ separation studies have indicated
that a pion fake rate of (0.07-0.5)\% can be achieved while maintaining
90\% electron identification efficiency for energies up to 40~GeV.
\end{abstract}

\end{frontmatter}

%\linenumbers

% main text
\section{Introduction}
\label{sec:segmentation}
The CERN Large Hadron Collider (LHC) is a 
proton-proton collider with 14~TeV centre of mass energy
and design luminosity of $10^{34}\mathrm{cm}^{-2}\mathrm{s}^{-1}$.
%Beam crossings are 25~ns apart and at design luminosity there
%are 23 interactions per crossing.
ATLAS is a LHC experiment designed to maximise the discovery
potential for new physics phenomena such as 
Higgs bosons and supersymmetric
particles, while keeping the capability of high-accuracy measurements
of known objects such as heavy quarks and gauge bosons.

%In the mass region between the LEP2 limit of 
%114.4~GeV~\cite{bib:lephiggs} and $2m_Z \sim 180$~GeV, one 
%of the most promising decay channel is $H\rightarrow \gamma\gamma$.
A very challenging mass region for Higgs discovery at the LHC lies 
between the limit reached by the LEP2 detectors at 114.4~GeV~\cite{bib:lephiggs} 
and $2\,m_Z \sim 180$~GeV. In this region, the decay $H\rightarrow \gamma\gamma$ is 
one of the most promising discovery channels. Its observation 
requires an excellent performance of the electromagnetic calorimeter in terms
of energy resolution, precision of angular measurement and particle identification capability. 

%the decay channel with the largest branching ratio, 
%$H\rightarrow b \bar{b}$,
%is very difficult to extract from the background, while channels with 
%clearer 
%signatures (for instance $H\rightarrow \gamma\gamma$) have much lower
%rates. At the time of the design of the EM calorimeter, the mass range
%114.4~GeV $<m_H<2m_Z$ was mainly covered by two
%decay modes, $H\rightarrow \gamma\gamma$ and $H\rightarrow 4\ell$. The
%possibility of observing the Higgs in this mass window through these 
%two channels set the requirements on the EM calorimeter
%performance in terms of energy resolution, precision of angular 
%measurement
%and particle identification capability. As an example, the signal/noise
%ratio for $H\rightarrow \gamma\gamma$ is 3-5\% in the mass bin of 
%interest. Since the intrinsic width of the Higgs is of a few MeV in 
%this mass region, the
%possibility of observing a signal peak in the $\gamma\gamma$ invariant 
%mass distribution over the background continuum depends crucially on 
%the energy
%and angular resolution of the EM calorimeter, and on the $\gamma$-jet
%separation. 
%\section{Calorimeter segmentation}

The electromagnetic calorimeter in the ATLAS experiment
is a lead-liquid argon sampling calorimeter with 
accordion shaped absorbers and electrodes. 
Liquid argon technology  has been chosen
because of its intrinsic linear behaviour as function of the deposited energy, 
stability of the response
and radiation tolerance.
\par
In the following, the beam direction defines the $z$ axis, and the
$x-y$ plane is transverse to the beam direction.
The azimuthal angle $\phi$ is measured around the beam axis, and
the polar angle $\theta$ is the angle from the beam axis; the
pseudorapidity is defined as $\eta=-\ln{\tan{\theta/2}}$.
The calorimeter has a cylindrical symmetry with a longitudinal segmentation
along the radius of the cylinder and transverse segmentation along the pseudorapidity
$\eta$. The central part covering $-1.5 < \eta < 1.5$ is housed in a separate
barrel cryostat. The two 
end-caps cover the region of $1.5 < | \eta | < 2.5 $.
 The transverse momentum $p_T$ is measured in the $x-y$ plane.
The optimisation of the longitudinal
and transverse segmentation 
of the calorimeter involves balancing performance issues such as 
electron and photon identification, position resolution, and pile-up and 
electronic noise contributions against cost and technical constraints such 
as routing of the signals from the calorimeter volume.
\par
\begin{figure}
  \centering
  \includegraphics[scale=0.65]{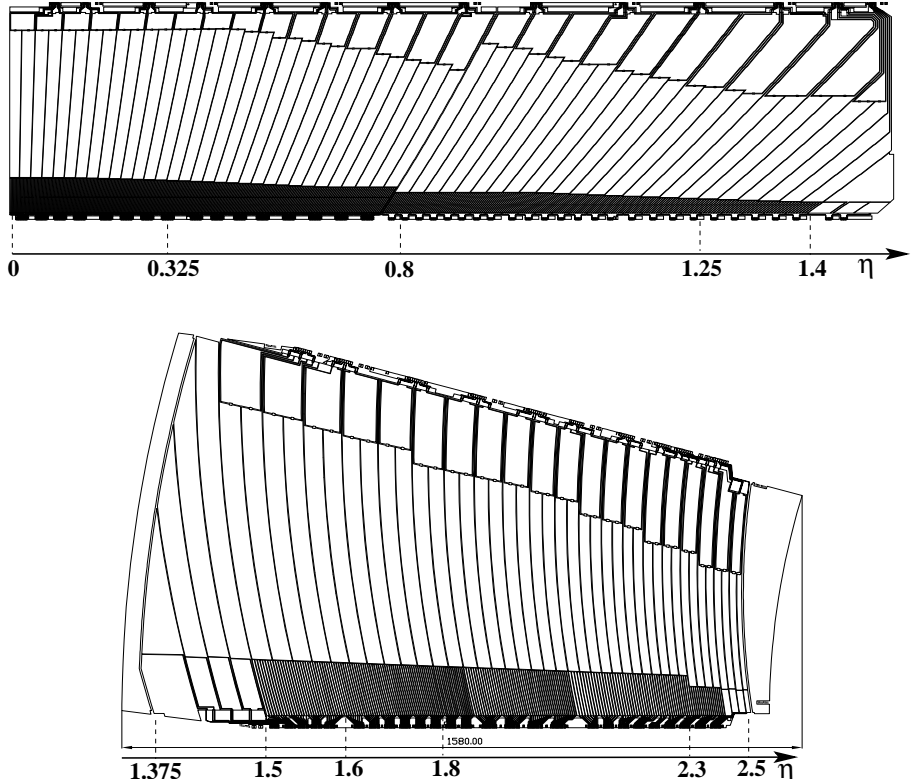}
  \caption{Segmentation of barrel and end-cap electrodes.}
  \label{fig:segmentation}
\end{figure}

The dominant backgrounds for the electrons and photons identification arise from the 
production
of hadronic jets.
%The most stringent physics requirement comes from 
%the search for the decay $H \rightarrow \gamma \gamma$
%which needs a very good rejection against single jets.
%which needs a rejection
%factor of about 5000 against single jets~\cite{bib:PTDR}.
The segmentations of the 
calorimeter~\cite{bib:PTDR}, sket\-ched 
in Fig.~\ref{fig:segmentation}, allow a
measurement of the shower shape that are an essential tool 
in background rejection. In addition, 
the rejection of spatially isolated signals from high-$p_T$ $\pi^0$'s coming 
from hadronic jet fragmentation is especially important:  ATLAS design goal calls for
a factor of three rejection of $\pi^0$ at a single-photon efficiency of 
90\% .
%In particular, a rejection of about three against
%the ones coming from single high-$p_T$ $\pi^0$'s from 
%jet fragmentation is
%needed. 
\par
At the LHC design luminosity, the beam spread in the $z$ coordinate
is about a few centimetres. The beam spread contributes to the angular
resolution of the the $\gamma\gamma$ invariant mass. In order to minimise
its effects, the photon direction has 
to be measured in $\eta$, with a 
resolution of about 50~mrad/$\sqrt{E(\mathrm{GeV})}$. In the $\phi$ direction, 
the interaction point is known with a precision of a few microns in the $x-y$
plane, and the azimuthal angle measurement is well constrained. 
All this 
requires a fine-grained position-sensitive device to perform angular
measurements and discriminate against the two merged
photons from the $\pi^0$ decay. This is done by segmenting the 
first
longitudinal sampling of the barrel (front compartment) 
into narrow cells of size $\Delta\eta \times
\Delta\phi \simeq 0.003 
\times 0.1$ . For the end-cap,  the segmentation of 
the first
sampling is $\Delta\eta \times \Delta\phi \simeq
0.003-0.006 \times 0.1$.
The other two compartments (middle and back) 
are segmented with towers of size
$\Delta\eta \times \Delta\phi = 0.025 \times 0.025\ (0.025 
\times 0.05)$ in
the second (third) sampling.
\par
In this paper the performances of barrel and end-cap modules, with
respect to the $\eta$ segmentation of their first compartment, are
presented. The results are based on analyses of test-beam data 
taken in 2000 with barrel module 0 (for the $\pi^0$ rejection calculation,
section~\ref{sec:gammapi0rej})
and  between 2001 and 2002 with series modules.

\section{Experimental setup}
A detailed description of barrel and end-cap modules of the ATLAS electromagnetic
calorimeter, and of the  signal reconstruction techniques can be 
found in Refs~\cite{bib:endcapnim,bib:barrelnim}. 
Two production modules, P13 for the barrel calorimeter and ECC1 for the end-cap were
tested in the CERN H6 and H8 beam lines during several months of tests during 2001-2002.
The test setup is similar to the one described in Refs~\cite{bib:endcapnim,bib:barrelnim}. 

\subsection{Electron run setup}
\begin{figure}
  \centering
  \includegraphics[scale=0.60]{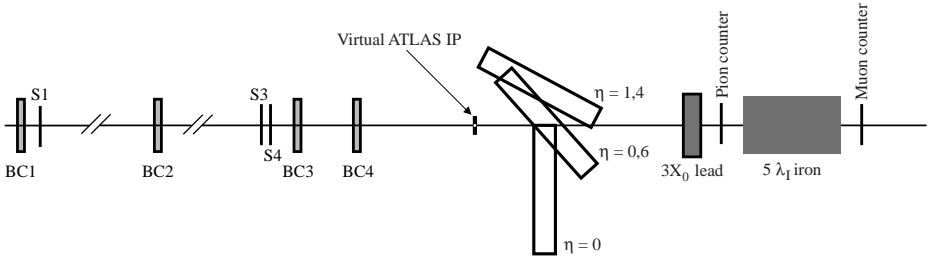}
  \caption{Electron beamline setup.}
  \label{fig:electron_beam}
\end{figure}

The calorimeter performance was tested using secondary or tertiary electron and pion beams, 
with momenta ranging from 20 to 245~GeV/c for barrel modules and from 20 to 150~GeV/c for 
end-cap modules.
The beam lines (Fig.~\ref{fig:electron_beam}) were equipped with three
scintillators (S1 to S3) in  
front of
the calorimeter for triggering purposes. Four multi-wire 
proportional
chambers%~\cite{bib:chambers} 
(BC1 to BC4) allowed to
determine the beam impact point at the calorimeter
with a resolution of about 250~$\mu$m in the $\eta$ direction. 
The size of
the last two scintillators, 
$4\times4~\mathrm{cm}^2$, defined the beam acceptance.
Cryostats housing the modules were mounted on remotely
controlled rails that allowed movements in $\eta$ and $\phi$ while 
ensuring incident angles similar to the ones expected in ATLAS for 
all positions. A 3$X_0$ lead absorber, 
a pion counter, a 5$\lambda_I$ iron absorber and a muon counter
were placed downstream of the cryostat.
$\eta$-scans were done at an electron energy of 245~GeV for the 
barrel and
120~GeV for the end-cap.
Energy scans at fixed positions were also carried out. 
%For the 
%barrel module, samples were taken at $\eta=0.69, \phi=0.28$~rad and
%energies in the range 20 to 245~GeV; for the end-cap module, 
%several samples were taken at $\eta=1.74, \phi=0.18$~rad and 
%energies in the range 20 to 150~GeV.

\subsection{Photon run setup}
\begin{figure}
  \centering
  \includegraphics[scale=0.60]{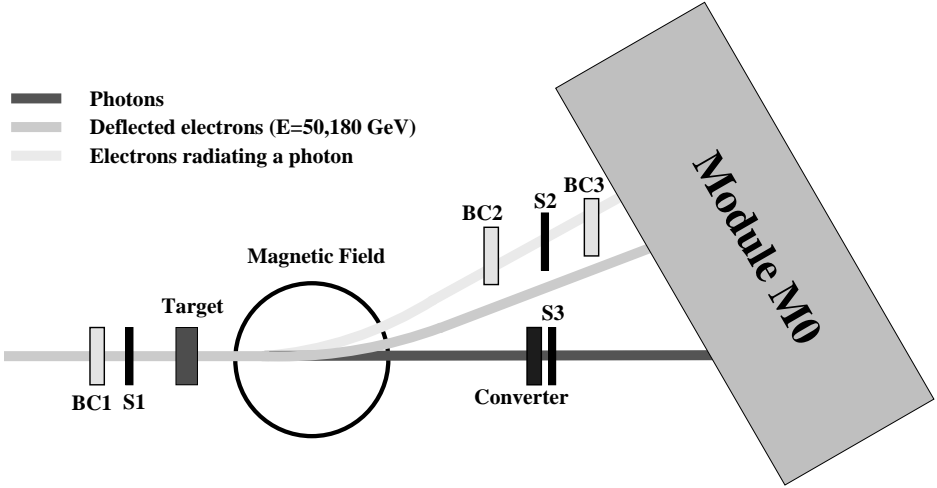}
  \caption{Photon beamline setup.}
  \label{fig:photon_beam}
\end{figure}

The calorimeter response to the incident photons was measured using
the prototype module M0. The beam line arrangement is illustrated in 
Fig.~\ref{fig:photon_beam}.
Electrons, with a momentum of 50 or 180~GeV/c impinged on a $0.1\ X_{0}$ 
lead target and produced bremsstrahlung photons. 
A 4~${\mathrm T} \cdot {\mathrm m}$ magnet, 
placed downstream of the target swept away the charged tracks from the
photon trajectory. The unaffected photon beam was incident on the module
at an angle and position corresponding to $\eta=0.69$ and 
$\phi=0.26$~rad. 
A beam chamber, BC1, was placed before the target and two more,
BC2-BC3, downstream of the magnet, in order to reject out of time counts.
%Three beam chambers, one before the target (BC1) and the other two 
%downstream of
%the magnet (BC2 and BC3), were used to reject accidental counts.
Beam trigger was provided by a coincidence of signal from two scintillator
counters S1 and S2. This second counter could be displaced in the
transverse direction with respect to the beam line in order to trigger 
on deflected electrons of different energy.
A $1 \ X_0$ thick converter, followed by scintillator S3
to detect photon conversions into $e^{+}e^{-}$ pairs, was placed in
the photon beam to enrich the data sample of events in which no more
than one photon was present; its effect is described more in detail in
section~\ref{sec:syst}.
The calorimeter was positioned 25~m downstream of the target.
%allowing a 
%separation between the photon beam
%and electrons which did not radiate of $D=\frac{3060}{E_{e}}$~cm, 
%$E_e$ being the electron energy in GeV. 
It was found that about $0.3 \ X_0$ of material was present along 
the beamline and acted as additional source of photons~\cite{bib:viret}.
In order to estimate the amount of events
for which more than one photon was present, a sample of photons 
produced with
the 180~GeV beam was also recorded without the photon converter.

\subsection{Pion run setup}
Positively charged pion beams of 20 and
40~GeV/c momentum, provided by the CERN H8 beamline, were used. 
The beam
impinged on series barrel module P15 of the calorimeter at fixed 
incident angle and position corresponding to $\eta=1.00$ and 
$\phi=0.28$~rad.
In addition to the usual test-beam setup, a  
\v{C}erenkov counter
was placed in the beam line and filled with He gas at the pressure of
$10^{-1}$~bar. The momentum threshold for a pion 
to generate
\v{C}erenkov light was 80~GeV/c and the counter provided 
an efficient discrimination 
of incident pions and electrons at 20 and 40~GeV/c beam momenta.

\section{Position and polar angle resolution}
\label{sec:posres}

The data were processed using the signal processing and pattern recognition
software EMTB~\cite{bib:emtb} to calculate the position and polar 
angle resolutions.
The event energy and
position were reconstructed using the
optimal filtering technique~\cite{bib:of,bib:lionel}, with a cluster size 
$\Delta \eta \times \Delta \phi$ of $24\times 1$ cells 
for the front and $3\times 3$ cells for the middle layer.
Random, muon- and pion-like events were discarded using trigger and 
scintillator information.
To correct for the energy loss upstream of the calorimeter,
the energy deposited in the presampler was
weighted by a factor $\alpha$ when calculating the total energy 
deposited
($E_{\mathrm{Cluster}}$). Longitudinal energy leakage induces 
a deterioration
of the energy resolution, therefore a weight $\beta$ was also 
applied to the energy deposited in the back sampling:
\begin{linenomath*}
\begin{equation}
E_{\mathrm{Cluster}} = \alpha E_{\mathrm{Presampler}} 
            + E_{\mathrm{Front}} +
  E_{\mathrm{Middle}} + \beta E_{\mathrm{Back}}
\end{equation}
\end{linenomath*}
The parameters $\alpha$ and $\beta$ were obtained by minimising 
the energy resolution for every $\eta$ position~\cite{bib:fanti}.

\subsection{Position resolution in the $\eta$ direction}
\label{sec:posresbar}

$\eta$-scans with cells at $\phi=0.26$~rad for the barrel and $\phi=0.18$~rad
for the end-cap were taken to study the variation of the position resolution 
in the $\eta$ direction.
%In the barrel, the $\eta$-scan with 
%cells at $\phi=0.26$~rad was considered.
Cells for which no optimal filtering coefficients were available 
were  
excluded from the analysis; the scanned region was therefore reduced
to
$0 \leq \eta \leq 1.14$ (barrel) and $1.50 \leq \eta \leq 2.40$ (end-cap). 
In addition, cells in the barrel at $\phi=0.28$~rad, 
$\eta=0.34,0.69$ and $0.96$, where a large number  
of events was accumulated, were used to 
evaluate the variation of the position resolution with respect to the 
impact point in the cell.
\par
\begin{figure}
  \centering
  \includegraphics[scale=0.5]{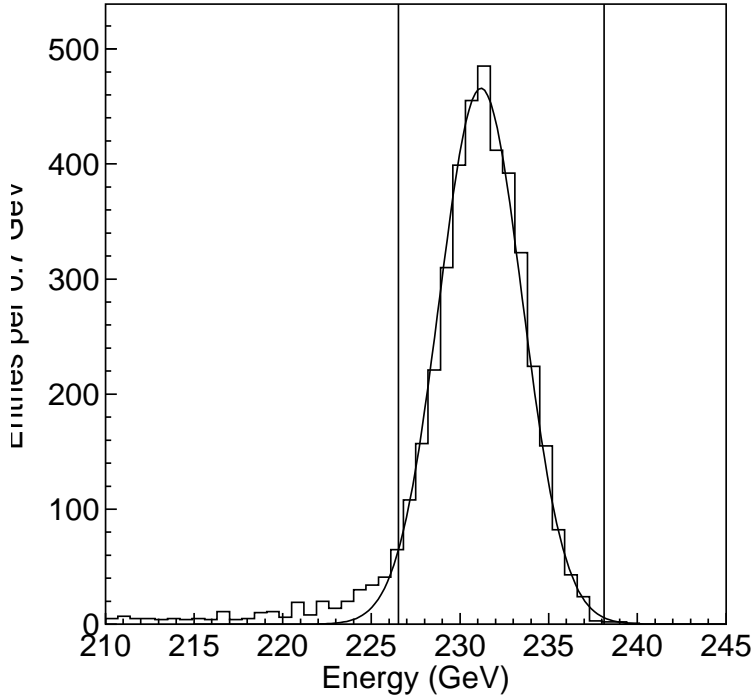}
  \caption{Typical energy distribution before the energy cut. The
    accepted
    window in the distribution is shown by the vertical lines.}
  \label{fig:enerfit}
\end{figure}

Pions in the beam are expected to deposit lower energy in the calorimeter
than electrons. In 
order to improve 
the pion rejection, it was required that the peak value of the energy distribution
fitted (Fig.~\ref{fig:enerfit}) to a Gaussian shape with peak value $E_{peak}$ and 
width $\sigma$ is in the range   
$-2<(E_{peak}-E)/\sigma<3$.
Additional constraints were imposed on the timing signal coming from 
the four beam chambers to guarantee a good track 
reconstruction and rejection of accidental background. 
Finally, the barycenters of the reconstructed showers in $\eta$ and $\phi$ 
were required to be within the central cell of a 3$\times$3 cluster 
in the middle compartment.
About 12\% of the events for the barrel module and 9\% 
of the
events for the end-cap module remained after all these requirements.
%at the end of the selection.
\par
The electron impact position of the beam particle on the calorimeter was 
calculated performing a linear fit through the points identified by the
beam chambers in each event. The resolution for each chamber,
obtained from the distribution of the residuals, was used to
estimate the uncertainty in the track extrapolation to the front face of 
the calorimeter.
Fitted tracks were extrapolated to the shower 
barycenter in both front and middle compartments. 
Except in the case where the electron track originated at the virtual ATLAS 
interaction point, the different depths of the shower barycenters and the 
electrodes' geometry resulted in different extrapolated values of $\eta$ in the front  
and middle compartments.
The chambers' resolution was estimated to be about 
250~$\mu$m (i.e. $1.5\times 10^{-4}$ in units of pseudorapidity) 
for both beamlines.
\par
Two methods were used to calculate the cluster
$\eta$ position in the calorimeter.

In the first method,
the position of an electron cluster in the samplings is the 
energy-weighted
average of the positions of the cells included in the cluster. 
Due to the steeply falling lateral profile of the EM shower, the 
finite
granularity of the calorimeter induces a systematic shift of the 
measured shower barycenter towards the centre of the cell. 
This effect is stronger in the $\eta$ direction, while 
the energy sharing is larger in the $\phi$ direction because of the 
accordion shape.
%In the case of the ATLAS accordion
%calorimeter, this is especially true in the $\eta$ direction, whereas
%in the $\phi$ view the accordion waves induce a better energy 
%sharing
%between neighbouring cells. 
In test-beam data, the
shower barycenter position measured by the calorimeter
can be compared to the 
position measured by the beam chambers.
The resulting distortions of the measured positions were fitted and used to
correct the barycenter position. The corrections depended on the beam energy and 
the $(\eta,\phi)$ direction. 
\par
The second method took into account the exponential 
transverse profile of the shower. Here the average of the positions of the 
cells included in the cluster was calculated by weighting
the individual contributions with the logarithm of the
cell energy, according to~\cite{bib:lw}:
\begin{linenomath*}
\begin{equation}
  \eta_C =  \frac{\sum_i w_i \eta_i}{\sum_i w_i},
\end{equation}
\end{linenomath*}
where the index $i$ runs over cells belonging to a given 
cluster in a compartment. In this case,
\begin{linenomath*}
\begin{equation}
  w_i = \max(0, w_0 + \ln(E_i/E_C)),
\label{eq:logw} 
\end{equation}
\end{linenomath*}
$E_C$ is the total energy deposited
in a given compartment, and $w_0$ is a parameter that 
defines a minimum fraction of the shower energy in the cell 
and sets the relative importance of the tails of the shower in the 
weighting procedure. With this reconstruction the position distortion 
practically disappears and $\eta_C$ can be directly 
compared with $\eta_{BC}$ without further corrections. 
\par
\begin{figure}
  \centering
  \includegraphics[scale=0.5]{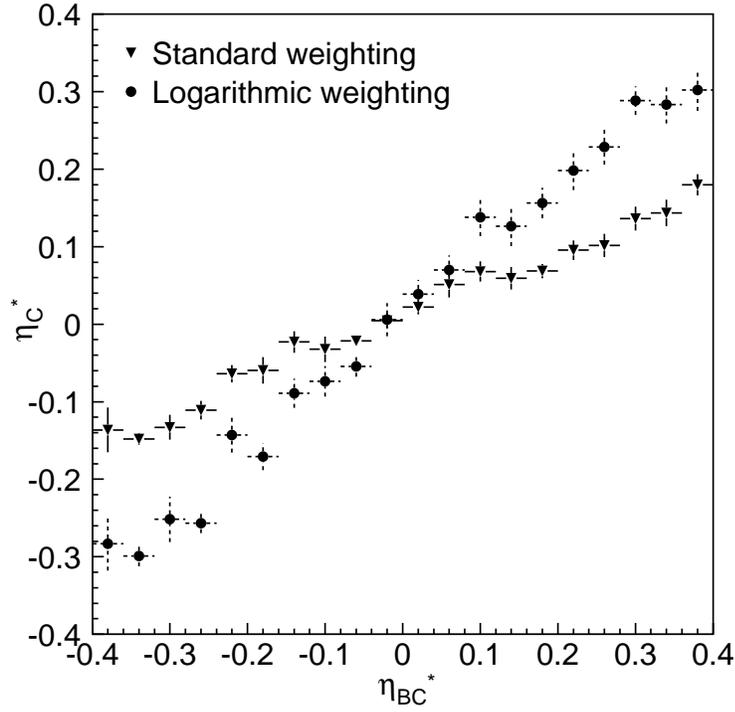}
  \caption{%Left: $\eta_C$ plotted against $\eta_{BC}$ for the barrel
    %at $\eta=0.44,\phi=0.26$, E=245~GeV. The dashed line is
    %a superposition of the fit function; the three points
    %corresponding
    %to the centre and borders of the cell lie on a straight line of
    %unit slope.
    %Right: 
    $\eta_C$ versus $\eta_{BC}$ distribution for standard and
    logarithmic weighting in the barrel front compartment at
    $\eta=0.69, \phi=0.28$~rad, $E=245$~GeV. The bias in
    $\eta_C$ disappears when using formula~\ref{eq:logw} for the
    weighting and
    the points are distributed along a line of unit slope.}
  \label{fig:sshape}
\end{figure}
\begin{figure}
 \centering
  \includegraphics[scale=0.65]{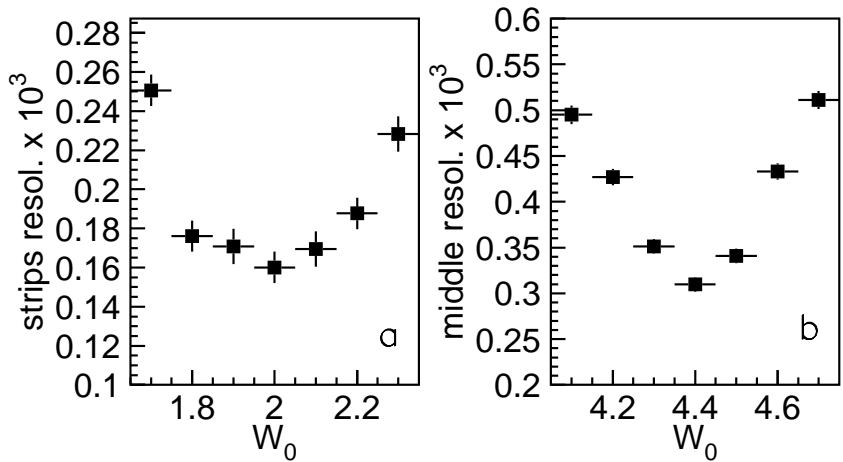}
  \caption{Variation of the $\eta$ resolution for the front (a) 
    and middle
    (b) compartment as a function 
    of the parameter $w_0$ at $\phi=0.26$~rad, $E=245$~GeV.}
  \label{fig:w0}
 \end{figure}

A comparison of 
$\eta_C$ versus $\eta_{BC}$ distributions for the cases of 
standard and
logarithmic weighting (LW) is presented in Fig.~\ref{fig:sshape}.
Here, the value of $w_0$ was 
chosen such that it gives a minimum for the position resolution. 
For the barrel module, the best value was
found to be 2.0 for the front and 4.4 for the middle compartment at
$\phi=0.26$~rad. For the end-cap module, at $\phi=0.18$~rad in the front
compartment, the value of $w_0$ was 2.2. 
%With the granularity of the front compartment, it may 
%happen that for certain $\phi$ positions the energy may be
%shared between two different $\phi$ cells. In such a case, 
%the amount of energy deposited 
%in each cell is smaller compared to the case when the cluster is
%contained in a single
%$\phi$ cell. This means that the weights to 
%be applied change accordingly; in particular, 
%the value of $w_0$ was found to be 2.3 at $\phi=0.28$~rad, for the front
%compartment of the barrel module.
A plot showing the variation of the resolution for the front and 
middle compartments  
as a function of $w_0$ at $\phi=0.26$~rad is presented in 
Fig.~\ref{fig:w0}.
%The weights need to be
%adjusted when the energy sharing between two cells along the $\phi$ 
%direction is large.

In the middle compartment the accumulated statistics per cell was
sufficient to perform a good fit of the position distortion. 
It was found that the
position correction done using LW for the middle compartment gave 
results worse by 20 to 60\%~\cite{bib:sacco} at 245~GeV.
\par
The case is different for the front compartment; its finer 
segmentation with respect to the middle compartment implied
a much smaller number of events populating a given strip.
The fits were therefore more difficult to perform and
only possible for the two strips nearest to the impact point
of the electron. For these strips it was found that the 
two weighting methods gave comparable results. In order to 
analyse the highest possible number of strips, 
it was decided to use the LW method to 
correct the reconstruction bias in the front compartment.

\subsubsection{Monte Carlo samples}
\label{sec:mcbarrel}
\begin{figure}
  \centering
  \includegraphics[scale=0.65]{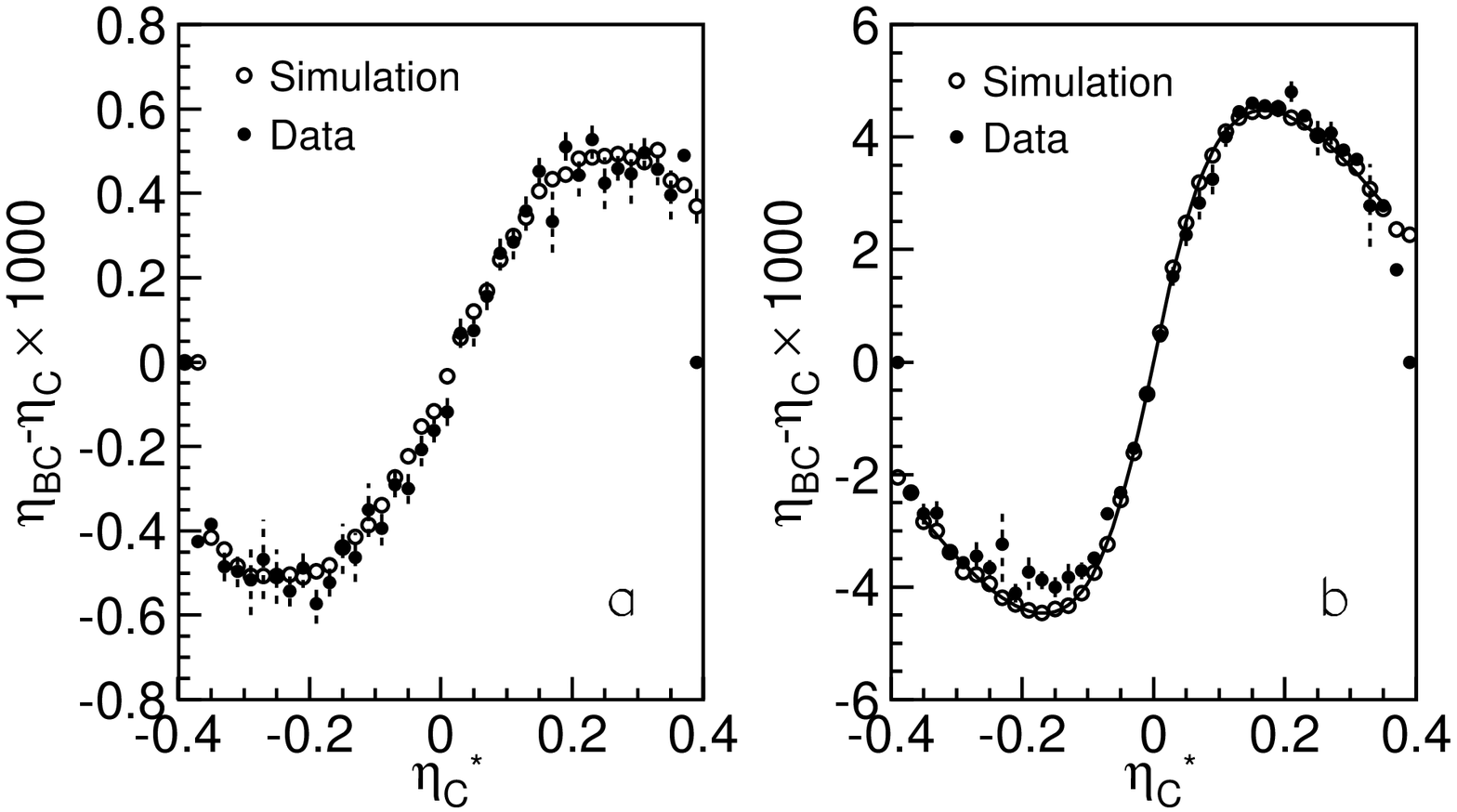}
  \includegraphics[scale=0.65]{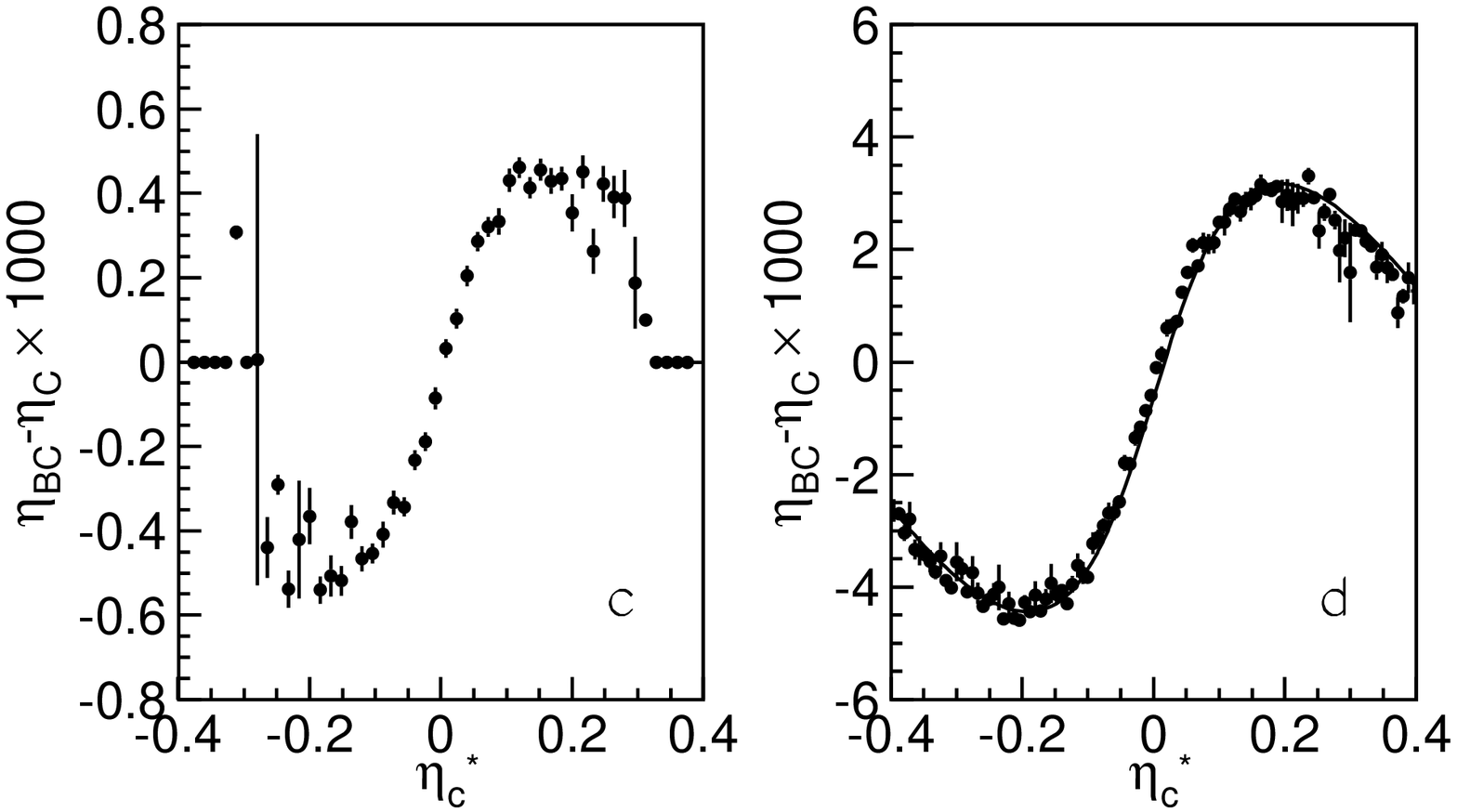}
  \caption{S-shape comparison between data and Monte Carlo for 
    front (a)
    and middle compartment (b) of a typical cell 
    of the barrel
    module, at $E=245$~GeV. The corresponding plots for the end-cap 
    module, at
    $E=150$~GeV, are shown in plots (c) and (d). The solid lines 
    represent the S-shape fit.} 
  \label{fig:sshcomp}
\end{figure}

GEANT3 simulation~\cite{bib:geant3}, adapted to reproduce ATLAS 
test-beam
events, was used to generate   
Monte Carlo samples for the barrel module. For each 
cell  $\phi =0.26$~rad,  
about 3000 events were generated. Three samples at $\phi = 0.28$~rad 
containing about 10000
events were also used to study cells for which a large 
number of
events was available at $\eta = 0.34,0.69$ and $0.96$.
The Monte Carlo samples were analysed with the EMTB package in which the 
clustering routine
was modified to take into account electronic noise and cross-talk. 
The cell noise level for presampler, front, middle and back 
compartments was
set to 45, 15, 30 and 25~MeV respectively. The cross-talk was
set to  4.1\% (1\%) between neighbouring front (middle)
compartment cells.
The comparison of the data and Monte Carlo is illustrated in
Fig.~\ref{fig:sshcomp} and shows a good agreement. 
%\par
%No Monte Carlo simulation could be produced for the end-cap module 
%in time for this analysis.

\subsubsection{Resolution calculation}
\label{sec:rescalc}
\begin{figure} 
  \centering
  \includegraphics[scale=0.65]{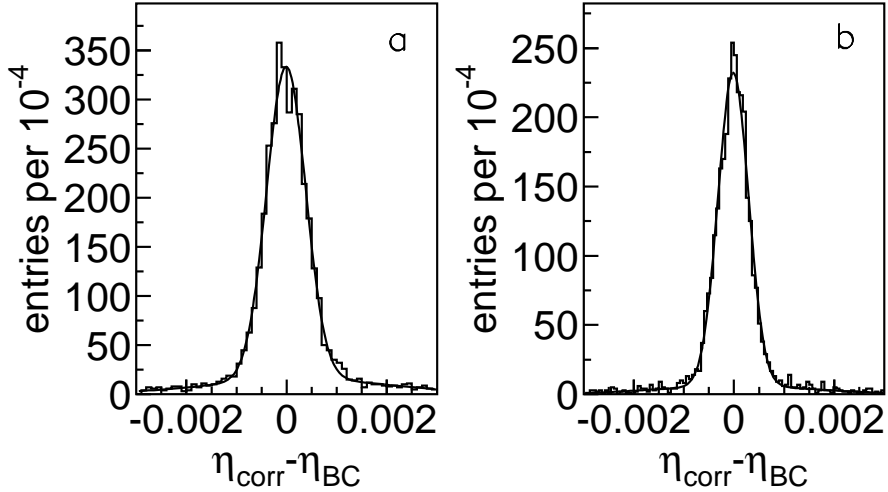}
  \caption{Double Gaussian fit to the
    $\eta_{\mathrm{corr}}-\eta_{\mathrm{BC}}$ distribution 
    for the barrel at 245~GeV (a) and the end-cap (b) at 150~GeV.
    The parameters are defined 
    in equation~\ref{eq:gg}.}
  \label{fig:gg}
\end{figure}

%Unfortunately, the beam chamber information lacked of an absolute 
%measurement
%of the incident beam direction with respect to the calorimeter 
%position. The
%data contained only an identification of 
%the cell hit by the electrons; it was impossible to know
%where across the cell the electron actually impinged. 
%This meant that the $\eta$ average values 
%calculated by the beam chambers were shifted with respect to those 
%measured by
%the module by a fraction of cell. 
%In order to overcome this difficulty, a constant value was added
%to $\eta_{BC}$ cell by cell so that its average value coincided 
%with $\eta_C$, 
%the latter calculated using the logarithmic weighting method in 
%order to avoid any bias introduced by the S-shape. 
For the middle compartment, the correction's shape was fitted to
the function: 
\begin{linenomath*}
\begin{equation}
S(\eta_C^*) = P_1 + P_2\ \eta_C^* + P_3 \arctan(P_4\ \eta_C^*),
\label{eq:sshape}
\end{equation}
\end{linenomath*}
where $\eta_C^*$ is $\eta_C$ normalised to the cell width, 
as illustrated 
in Fig.~\ref{fig:sshcomp}.
For the barrel, the fit was performed to the Monte Carlo samples 
and the resulting correction 
applied directly to the data. For the end-cap, the fit was 
performed directly to the data. The corrected value for the barycenter 
was then compared with the one obtained by track extrapolation. 
The resulting 
$\eta_{\mathrm{corr}} - \eta_{\mathrm{BC}}$ distribution was fitted 
with a sum of two 
Gaussians, in order to take into account the non-Gaussian tails due to 
residual accidental hits in the beam chambers,  
as shown in Fig.~\ref{fig:gg}. The parameters of the two 
Gaussians were defined by
\begin{linenomath*}
\begin{equation}
f(x)=P_1\ e^{{\textstyle -\frac{1}{2}(\frac{x-P_2}{P_3})^2}}+
        P_4\ e^{{\textstyle -\frac{1}{2}(\frac{x-P_5}{P_6})^2}}.
\label{eq:gg}
\end{equation}
\end{linenomath*}

\begin{figure}
  \centering
  \includegraphics[scale=0.48]{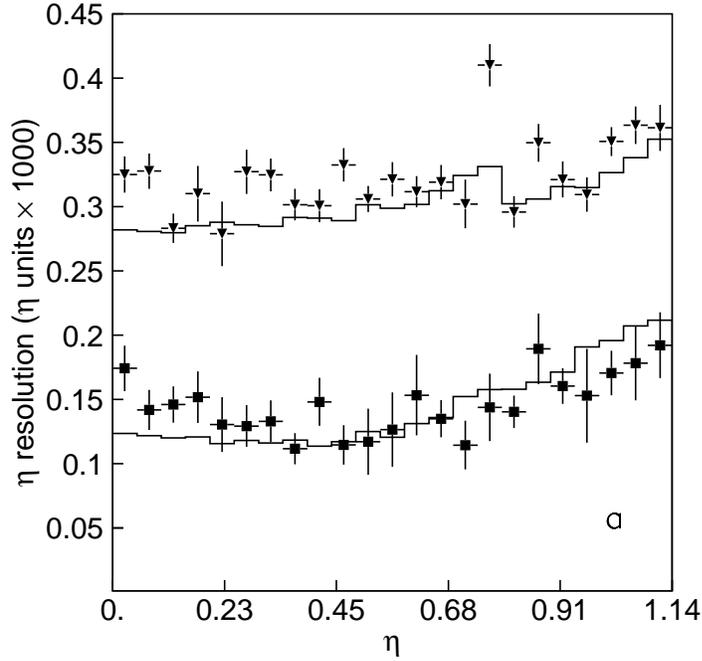}
  \includegraphics[scale=0.48]{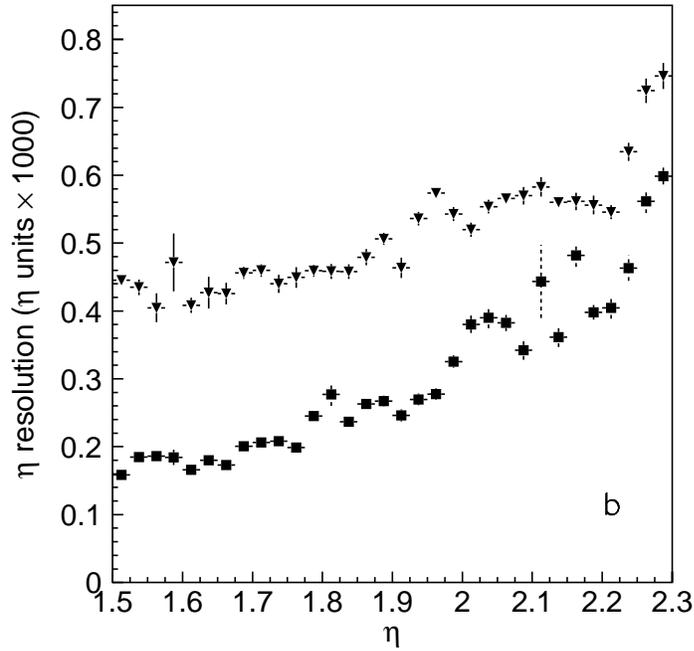}

 \caption{(a) $\eta$ resolution as function of $\eta$, 
    at $\phi=0.26$~rad, $E=245$~GeV for the front (squares) and middle 
    compartments (triangles) of barrel module P13, superimposed on the
    Monte Carlo prediction.(b) $\eta$ resolution as function
    of $\eta$,  
    at $\phi=0.18$~rad, $E=150$~GeV for the front (squares) and middle 
    (triangles) compartments of the end-cap module ECC1.}
  \label{fig:newplot}
\end{figure}

The resolution
of the beam chambers was subtracted in quadrature. The results 
for the middle and front compartments are summarised
in Fig.~\ref{fig:newplot} for barrel and end-cap modules.
In the barrel, the position resolution for both 
compartments does not have strong variation with $\eta$ and agrees well
with Monte Carlo simulation. The finer granularity of the 
front compartment results in a better position resolution.
Using the value of the resolution at small $\eta$ for the 
barrel and the appropriate  
extrapolation distance for the two compartments, it was possible 
to estimate the resolution as $240\ \mu$m for the front and $540\ \mu$m 
for the middle compartment.
In the end-cap, the worsening of the resolution at large $\eta$ is
expected and is due purely to geometrical effects.
%the geometrical increase in the dead material.

\subsubsection{Resolution variation within a cell}
\label{sec:histat}
\begin{figure}
 \centering
  \includegraphics[scale=0.4]{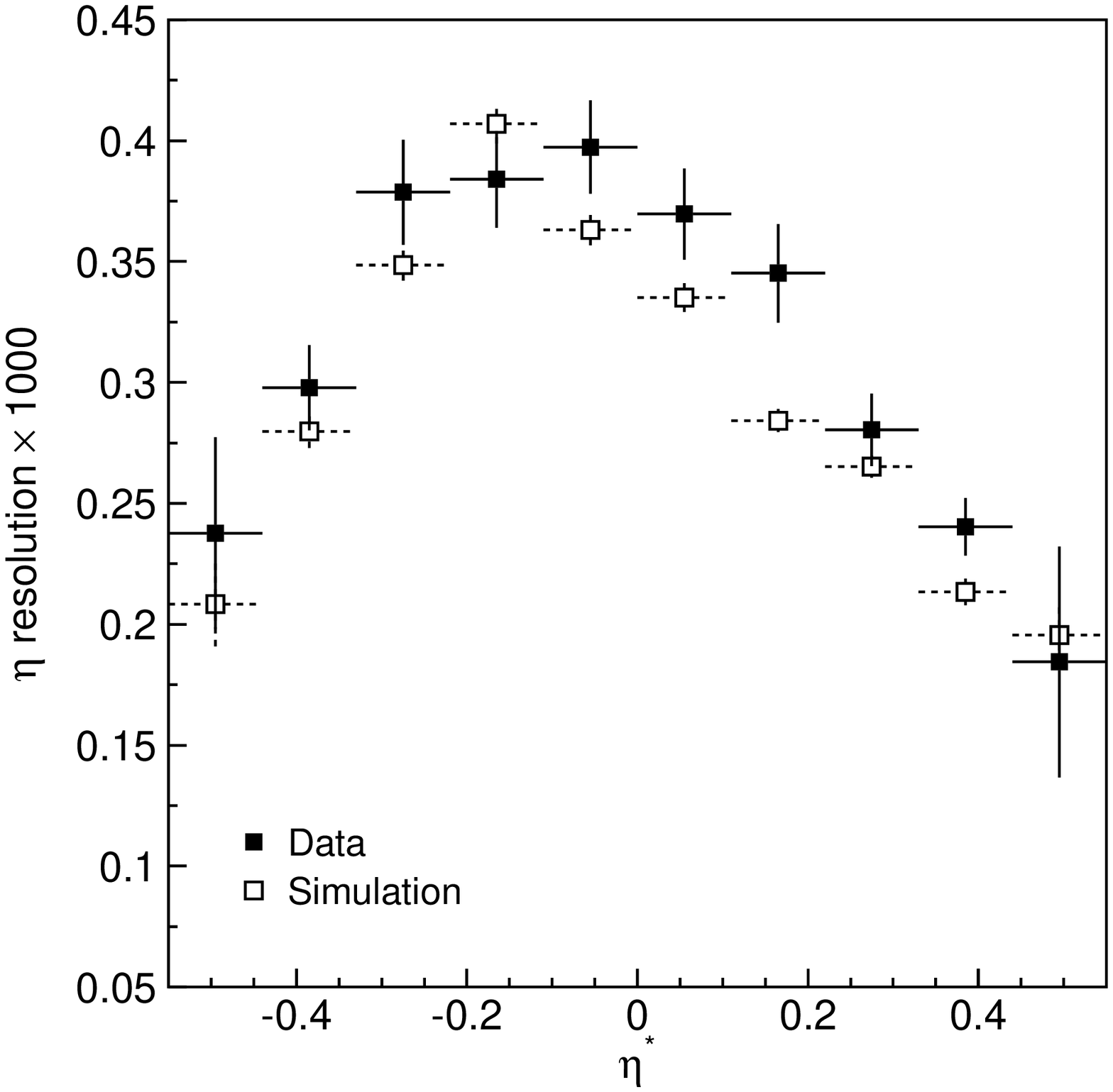}
  \caption{Variation of $\eta$ resolution across cell 
    $\eta=0.69, \phi=0.28$~rad, for the barrel module P13, at $E=245$~GeV. 
   % Right: variation of $\eta$ resolution within a cell for end-cap 
   % module ECC1 {\bf To be added.}.
  }
  \label{fig:incell13}
 \end{figure}

Using the runs for which a large number of events was accumulated, 
it was
possible to check how the resolution varied within a single cell of 
the middle compartment. 
After obtaining the correction from a fit to the Monte 
Carlo sample, the cell 
was divided into 10 slices in $\eta$ and the correction applied. 
Again, the 
$\eta_{\mathrm{corr}} - \eta_{\mathrm{BC}}$ distribution was fitted 
for each slice to a sum of two Gaussians, 
and the width of the narrower Gaussian was compared to the prediction 
of Monte Carlo.
The variation of the resolution across a single cell is shown in 
Fig.~\ref{fig:incell13} 
for the barrel module at $\eta=0.69, \phi=0.26$~rad. As expected, 
the resolution
is better at the border of the cell because of the better energy 
sharing
among cells in the cluster; the lopsided symmetry is due to 
the geometric shape of the cell and the variation of the
amount of material along $\eta$.
%The plots for 
%the end-cap modules, with no Monte Carlo comparison, are also 
%presented. {\bf To be added.}  

\subsubsection{Resolution energy dependence}
\begin{figure}
  \centering
  \includegraphics[scale=0.3]{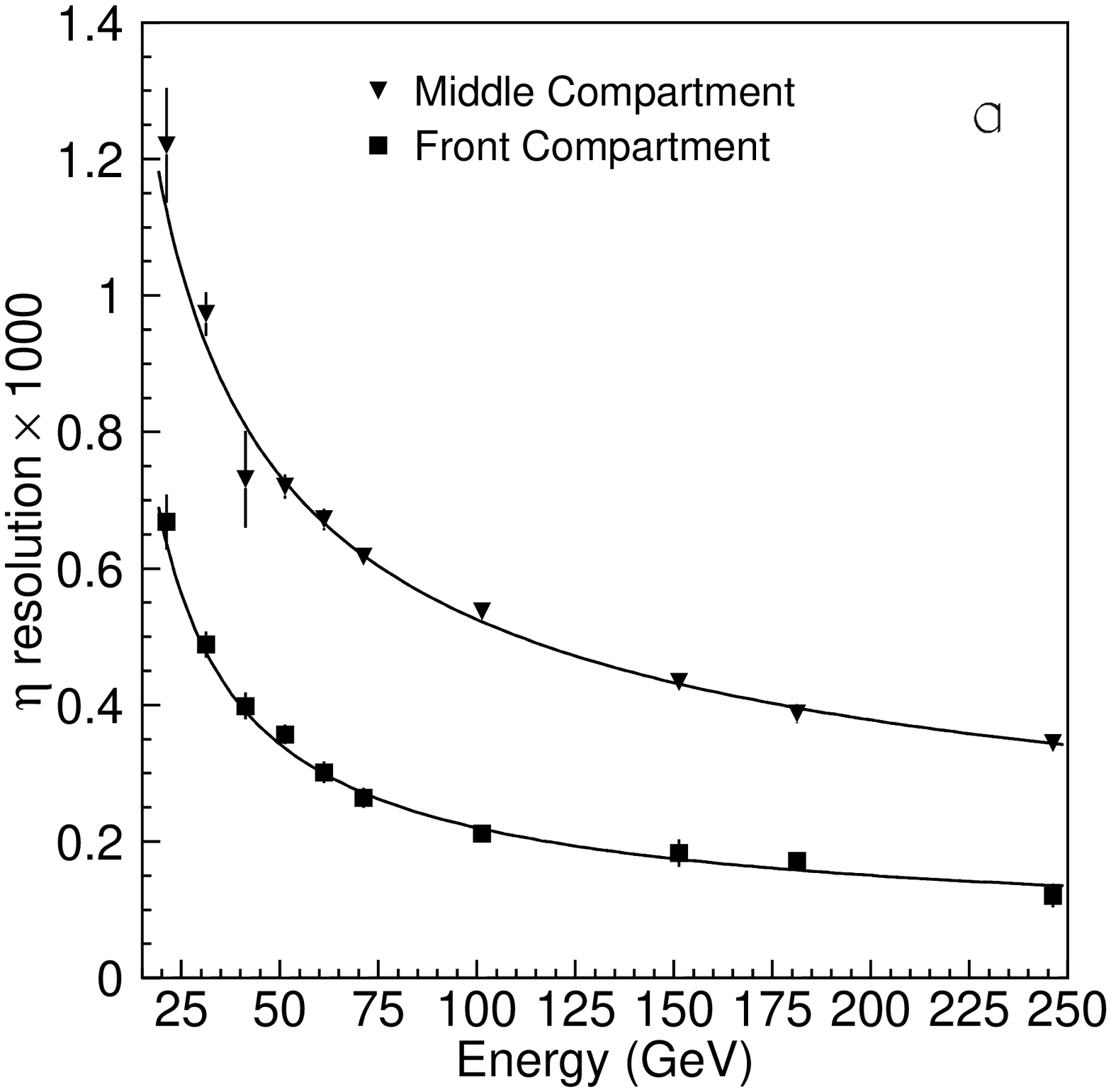}
  \includegraphics[scale=0.3]{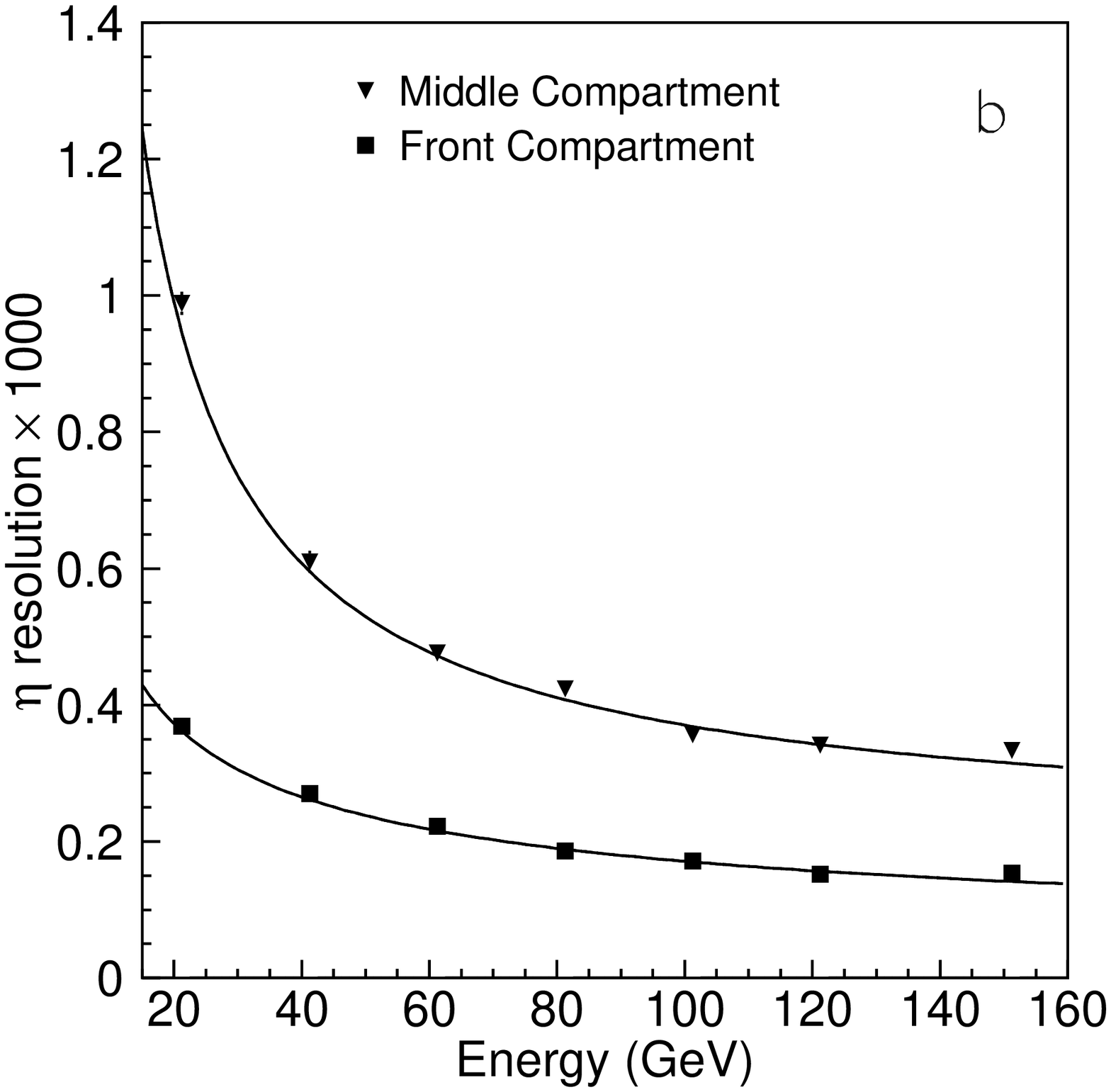}
  \caption{(a) $\eta$ resolution as function of the beam energy 
    for the middle (triangles) and front (squares) 
    compartment of barrel module
    P13 at $\eta=0.69, \phi=0.28$~rad; the results of the fit are 
    presented in Table~\ref{tab:resener}. (b) $\eta$ resolution 
    as function of the beam energy,  
    at $\eta=1.74, \phi=0.18$~rad for the middle (triangles) and front
    (squares) compartments of end-cap module 
    ECC1. The results of the fit are 
    presented in Table~\ref{tab:resener-ec}.}
  \label{fig:resener}
\end{figure}

For the barrel module, several data samples were studied for a position 
$\eta=0.69, \phi=0.28$~rad and at several different beam  
energies in the range of 20~GeV to 245~GeV. 
The $\eta$ resolution was calculated as
described in section~\ref{sec:rescalc}, and the resulting variation
of the resolution for different beam energy was fitted with a function:
\begin{linenomath*}
\begin{equation}
\sigma(E) = C_1 \oplus \frac{C_2}{\sqrt{E}} \oplus \frac{C_3}{E},
\label{eq:resener}
\end{equation}
\end{linenomath*}
where $E$ is the beam energy. The result of the fit is presented in 
Fig.~\ref{fig:resener} and Table~\ref{tab:resener}.
For the end-cap module, several samples were studied at 
$\eta=1.74, \phi=0.18$~rad and different 
energies in the range of 20~GeV to 150~GeV. The results of the fit are 
shown in 
Fig.~\ref{fig:resener} and Table~\ref{tab:resener-ec}.

\begin{table}[t!]
  \centering
  \begin{tabular}{cccc}\hline\hline
    & $C_1$ & $C_2~(\sqrt{\mathrm{GeV}})$ & $C_3$~(GeV)  \\
    \hline
    Strips & $(0.40\pm0.88)\times 10^{-4}$ 
             & $(1.91\pm0.22)\times 10^{-3}$
               & $(1.03\pm0.13)\times 10^{-2}$ \\
    Middle & $(1.20\pm0.50)\times 10^{-4}$ 
             & $(5.05\pm0.20)\times 10^{-3}$
               & $(0.65\pm1.14)\times 10^{-2}$\\
    \hline\hline
  \end{tabular}
  \caption{Parameters defined in equation~\ref{eq:resener} 
    for front and middle
    compartment of barrel module P13.} 
  \label{tab:resener}
\end{table}

\begin{table}[t!]
  \centering
  \begin{tabular}{cccc}\hline\hline
    & $C_1$ & $C_2~(\sqrt{\mathrm{GeV}})$ & $C_3$~(GeV)  \\
    \hline
    Strips & $(0.43\pm0.16)\times 10^{-4}$ 
             & $(0.17\pm0.03)\times 10^{-2}$
               & $(0.17\pm23)\times 10^{-4}$\\
    Middle & $(0.19\pm0.03)\times 10^{-3}$ 
             & $(0.27\pm0.03)\times 10^{-2}$
               & $(1.15\pm0.10)\times 10^{-2}$ \\
   \hline\hline
  \end{tabular}
  \caption{Parameters defined in equation~\ref{eq:resener} for 
    front and middle compartment of end-cap module ECC1.} 
  \label{tab:resener-ec}
\end{table}

\subsection{Polar angle resolution}
\label{sec:polangresbar}
\begin{figure} 
  \centering
  \includegraphics[scale=0.6]{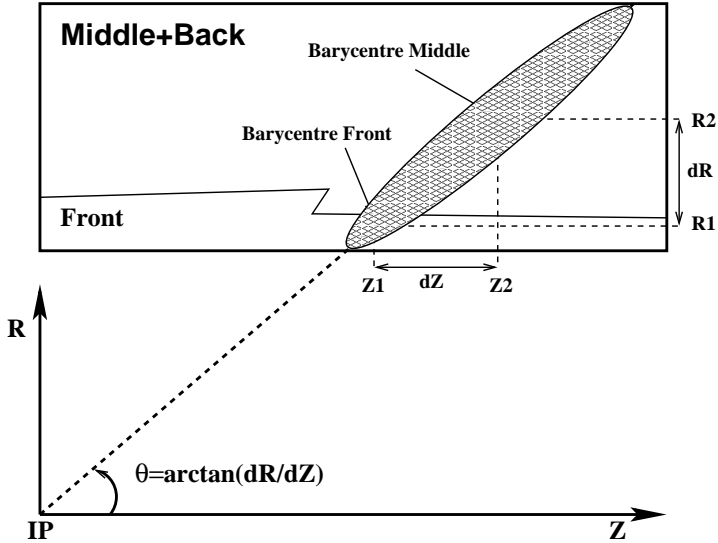}
  \caption{Schematic drawing of a barrel module in the $Z,R$ 
    plane and definitions of related quantities.}
  \label{fig:thetaplot}
\end{figure}

\begin{figure} 
  \centering
  \includegraphics[scale=0.33]{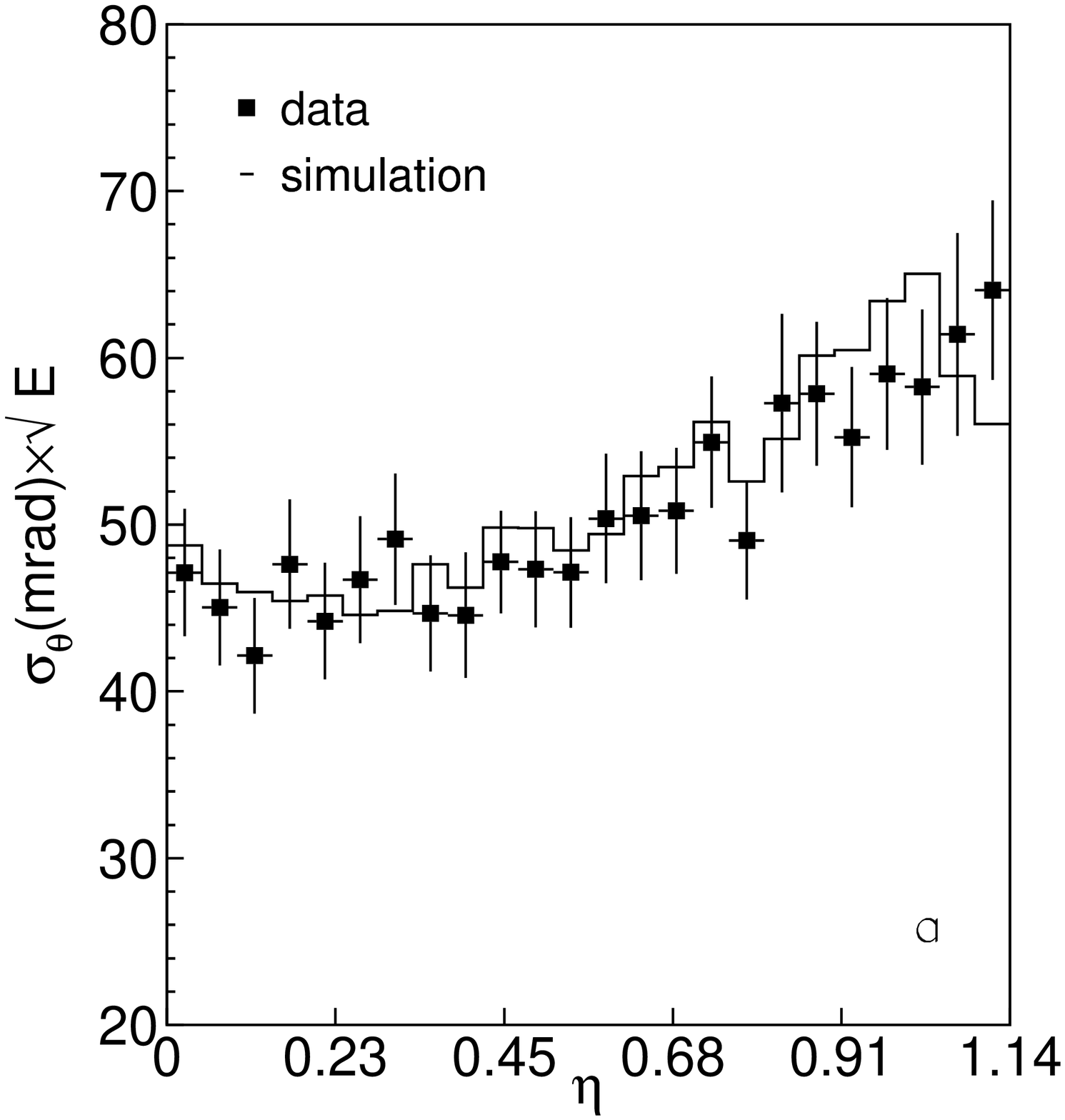}
  \includegraphics[scale=0.33]{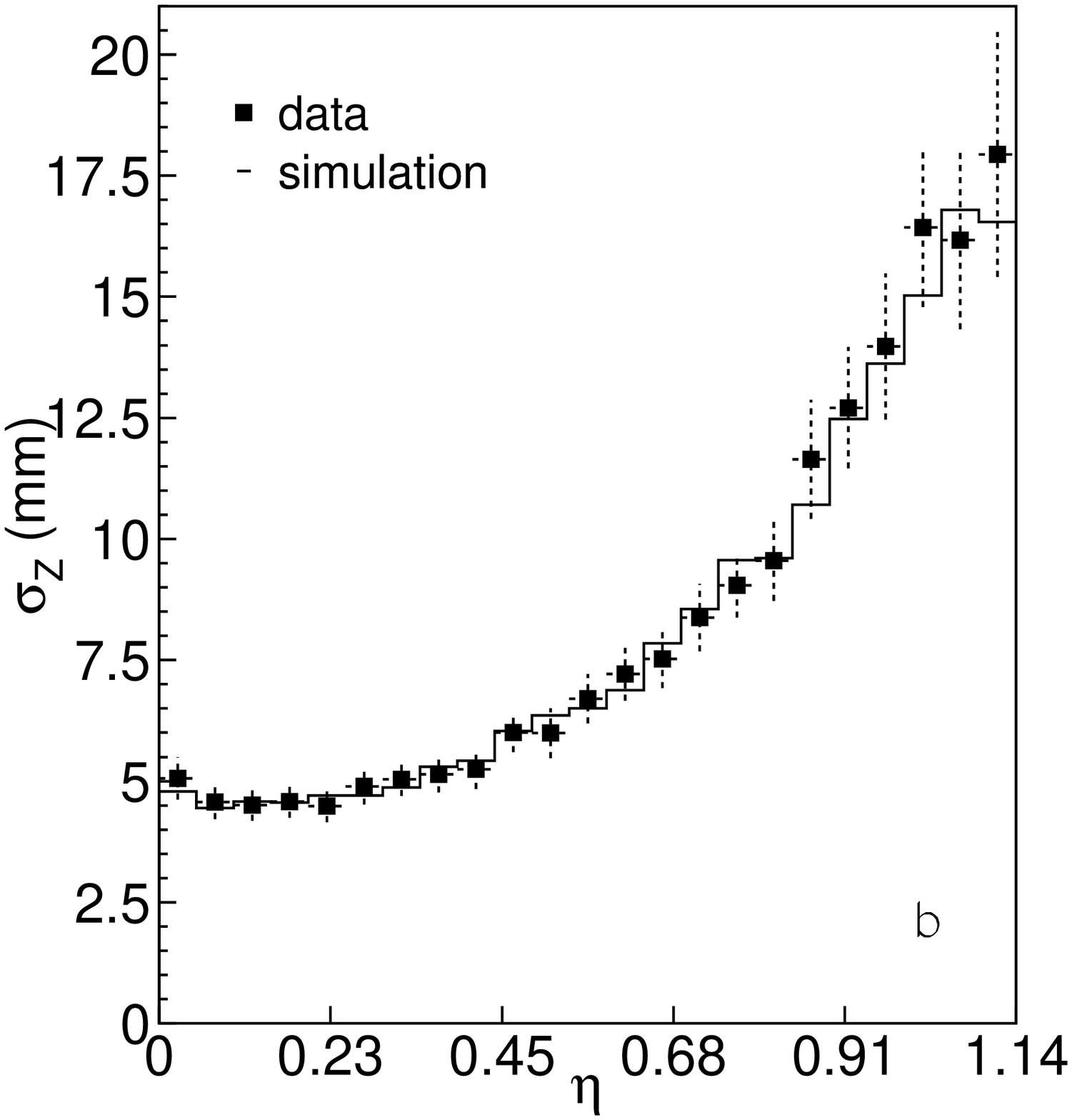}
  \caption{Angular resolution in  $\theta$ times the square root of the beam 
    energy
    (a) and $z_{\mathrm{vertex}}$ resolution (b) 
    as function of $\eta$ for barrel module P13,
    obtained using information from the front and middle 
    compartments
    at $\phi=0.26$~rad, at $E=245$~GeV.}
  \label{fig:zeta}
\end{figure}

\begin{figure} 
  \centering
  \includegraphics[scale=0.33]{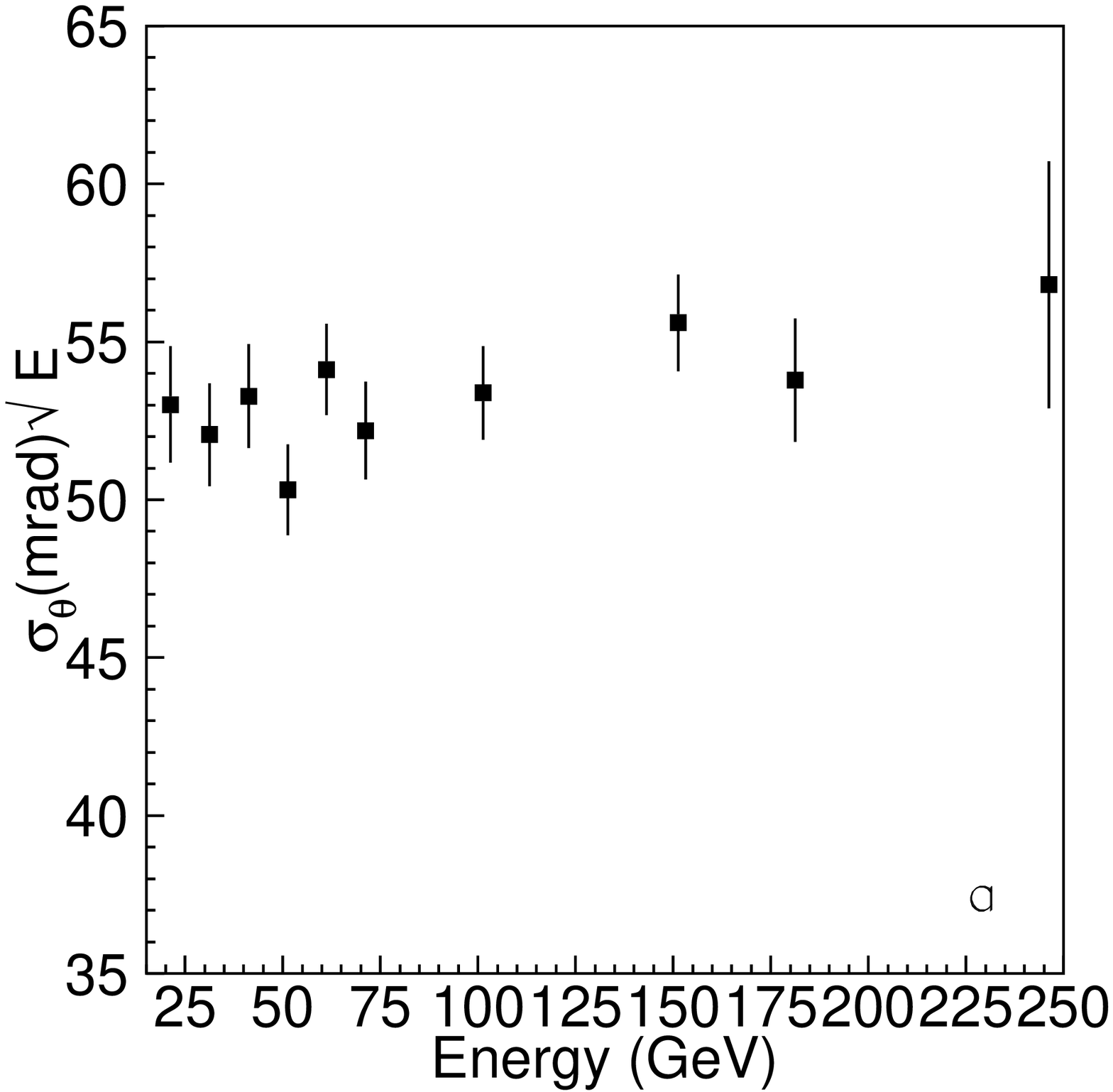}
  \includegraphics[scale=0.33]{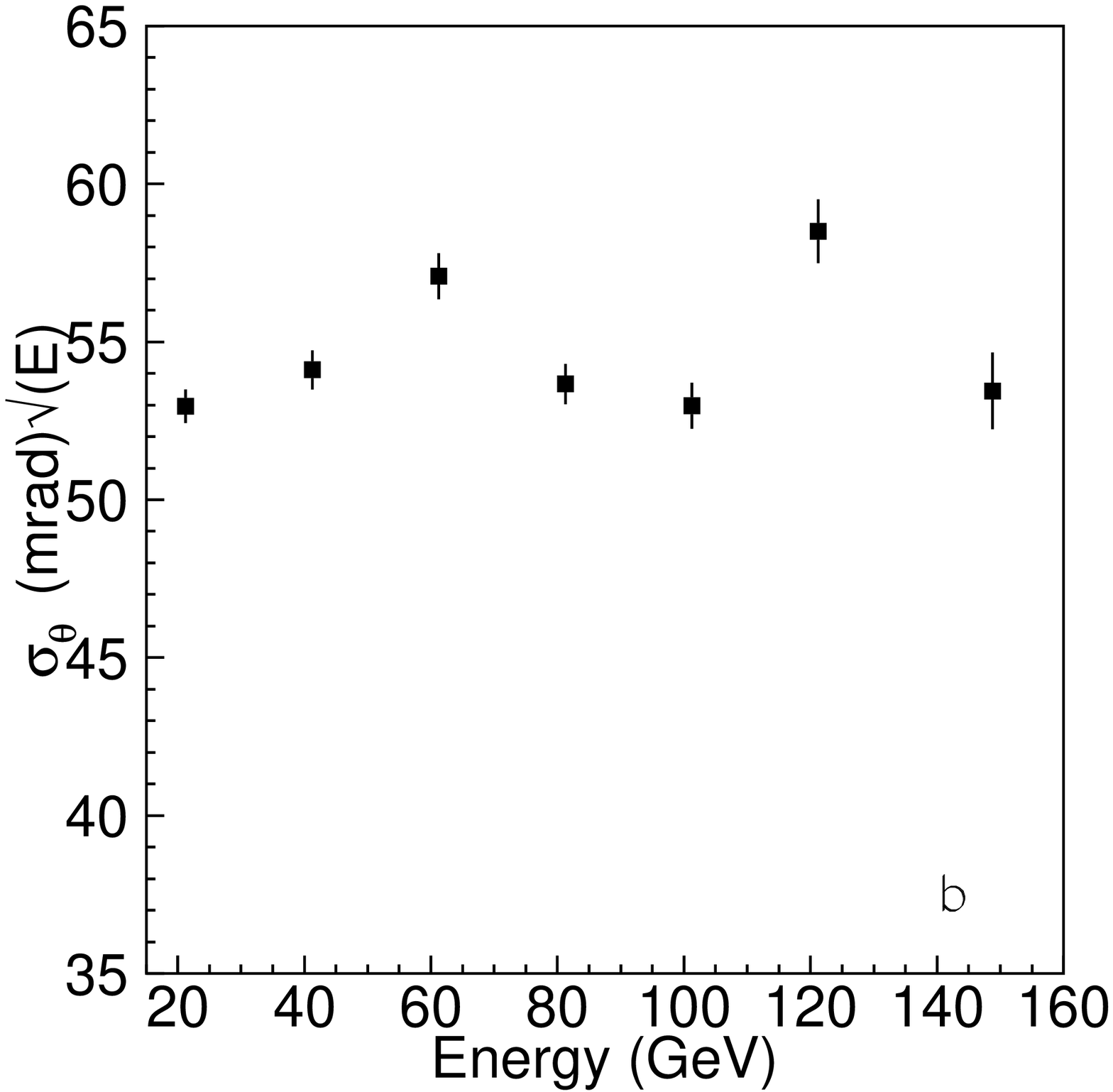}
  \caption{Variation of the polar angle resolution with energy for 
    barrel at
    $\eta=0.69, \phi=0.28$~rad (a)
    and end-cap at $\eta=1.74, \phi=0.18$~rad (b).}
  \label{fig:theta}
\end{figure}

The knowledge of the shower barycenters in the first two compartments
of the calorimeter allowed an extraction of the 
vertex position and the direction of the incident electron, 
relying only on calorimeter information. These values can be compared 
with values measured by the beam chambers (for the data) or
generated (for the simulation). 
The $(Z,R)$ reference frame used in the following is described in
Fig.~\ref{fig:thetaplot}. 
Here, $Z$ is the coordinate along the
ATLAS beam line and $R$ is the radial coordinate with respect to the 
same beam line, found for each cell in $\eta$ using the 
extrapolation distance to the
shower barycenter. % defined in section~\ref{sec:sshape}. 
The vertex position in $Z$ is given by:
\begin{linenomath*}
\begin{equation}
\frac{Z_2-Z_{\mathrm{vertex}}}{Z_2-Z_1}=\frac{R_2}{R_2-R_1} \quad
\Rightarrow \quad
Z_{\mathrm{vertex}}=\frac{Z_1 R_2-Z_2 R_1}{R_2-R_1}.
\end{equation}
\end{linenomath*}
The direction of the incident electron can be calculated as 
\begin{linenomath*}
\begin{equation}
\theta = \arctan{\frac{Z_2 \tan{\theta_2} - Z_1 \tan{\theta_1}}
                                                     {Z_2 - Z_1}},
\end{equation}
\end{linenomath*}
where $\theta_i$ are the angles pointing to the barycenter for the front 
and middle compartment respectively.
The polar angle resolutions were 
also 
calculated as function of the beam energy using the energy scan at 
$\eta=0.69, \phi=0.28$~rad for the barrel and
$\eta=1.74, \phi=0.18$~rad for the end-cap.
The resulting resolutions are plotted 
in Figs.~\ref{fig:zeta} and~\ref{fig:theta}.
%\begin{figure} 
%  \centering
%  \includegraphics[scale=0.4]{rzeta-1-13-7.eps}
%  \caption{$z_{\mathrm{vertex}}$ resolution 
%    as function of the energy, for end-cap module ECC1,
%    calculated using information from the front and middle 
%    compartments
%    at $\eta=1.74, \phi=0.18$~rad.}
%  \label{fig:zetares-ec}
%\end{figure}
For both barrel and end-cap, the measured polar angle resolution 
is in the
range 50-60~mrad/$\sqrt{E(\mathrm{GeV})}$, in good agreement 
with the design expectations.

%%%%%%

\section{$\pi^0$ rejection}
\label{sec:gammapi0rej}
%In the following analysis, it is assumed that the pre-selection 
%efficiency
%(i.e. based on hadronic calorimeter requirements, energy deposit on
%middle sampling, etc.) is identical for single photons and neutral
%pions.

The $\pi^0$ rejection is defined as
\begin{linenomath*}
\begin{equation}
R = \frac{1}{\epsilon_{\pi^0}} = \frac{P_0}{P},
\end{equation}
\end{linenomath*}
at fixed photon selection efficiency,
where $\epsilon_{\pi^0}$ is the pion selection
efficiency, $P_0$ and $P$ the number of neutral pions before and after the 
selection requirements. In the present analysis, the requirements were chosen
such that the photon selection efficiency on a $p_T=50$~GeV/c photon sample 
was 90\%. As no $\pi^0$ beam was available, a $\pi^0$-like sample has been 
produced by superimposing photons according to the kinematics of the 
$\pi^0$ decay. 

\subsection{Event selection}
\label{sec:evsel}
\begin{figure} 
  \centering
  \includegraphics[scale=0.65]{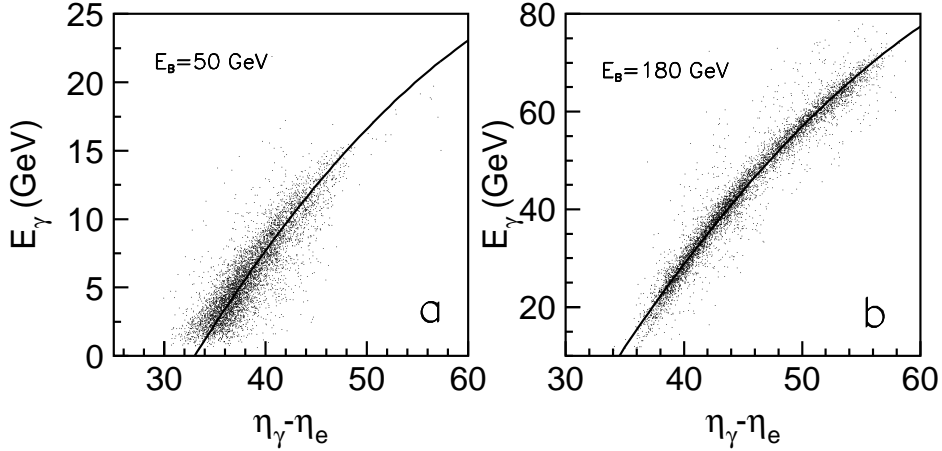}
  \caption{Photon energy versus distance between photon and electron
  clusters, in strip units, for beam energy of 50~GeV (a) and 
  180~GeV (b). The solid lines represent the expected behaviour.}
  \label{fig:cleanup}
\end{figure}

In order to have a sample of photons as clean as possible, only runs
taken with the converter placed in the photon beam were used to 
compute the $\pi^0$ rejection. 
To allow tagging, the 
electron had to hit the sensitive area of the last two beam chambers.
The rejection of accidental background and events in which the 
electron showered
before reaching the calorimeter was ensured by the requirement 
that the signal
coming from the beam chambers' TDCs be consistent with the passage 
of a single electron. About 30\% of the events satisfied all
these requirements. To avoid shower overlap, the electron and photon
positions had to be separated by 
more than 3 cells in the middle compartment (i.e. $\Delta\eta > 0.075$).
The sum of the photon and electron energies had to be consistent 
with the
beam energy, thus allowing the rejection of events in which the 
photon was
lost in the collimator. All these requirements reduced the sample to 5\% of 
its initial size.
The rejection of most multi-photon events (i.e. events for which more
than one photon was present) was
done using as photon veto the scintillator S3 located downstream of the
photon converter. 
In fig.~\ref{fig:cleanup} is shown the photon energy as function of the 
distance between
the photon and electron clusters for the two beam energies. 
Events more than $2.5\sigma$ away from the expected distribution 
were rejected.
%Events for which
%entries in this plot 
%did not agree at $2.5\sigma$ level with the expected distribution 
%were
%rejected. 
This last requirement reduced the sample by an additional factor of 
3. The energy for each strip was corrected to take into account 
the 4.1\% cross-talk described in section~\ref{sec:mcbarrel}. 
Finally, further multi-photon event rejection was achieved by 
requiring the
height of the second maximum in the strip energy distribution to 
be less than
200~MeV, value determined from Monte Carlo simulation (see
section~\ref{sec:tools}). After all cuts, about 6450 photon 
events were kept in the sample. 
\par
In order to calculate the $\pi^0$ rejection, pairs of test-beam
photons were used to build 
a $\pi^0 \rightarrow \gamma \gamma$ decay, modelled according to a 
simple kinematic simulation at $p_T=50$~GeV/c and $\eta=0.69$. The simulation 
provided a set of kinematic configurations, i.e., photon momenta in the
laboratory reference frame. The selection was performed in terms of energy
and distance between the two photons in the $\eta$ direction. 
The energies
had to match within $\pm 2$~GeV with the values expected from the 
kinematic simulation. Their 
relative distance had to be within 0.03~strip widths in $\eta$. 
After this selection, the $\pi^0$-like sample was made by 
superimposing the two photons: their energy deposit was summed  
strip by strip. 

\subsection{Monte Carlo samples}
GEANT3 simulation~\cite{bib:geant3}, adapted to reproduce ATLAS 
test-beam
events, was used to generate a 
Monte Carlo photon sample, consisting of
about 8000 events, at $\phi = 0.26$~rad, $\eta=0.69$ and in the 
energy range
0-70~GeV. A sample of direct $\pi^0
\rightarrow \gamma \gamma$ decays was also generated at 
$\phi = 0.26$~rad,
$\eta=0.69, p_T=50$~GeV/c to
check the consistency 
of the method described above.
\par
These Monte Carlo
samples were analysed with the EMTB package as described in
section~\ref{sec:mcbarrel}. 
%in which the clustering routine
%was modified to take into account electronic noise (see
%section~\ref{sec:mcbarrel} for details). 
The cross-talk
between strips was corrected for in the data, but a consistency 
check was
carried out by introducing cross-talk in the simulated samples 
instead; the
resulting agreement 
between data and Monte Carlo for the $\gamma/\pi^0$ separation was 
unchanged
with respect to the standard analysis.

\subsection{$\pi^0$ rejection calculation}
\label{sec:tools}
Photons and neutral pions can be separated by analysing the shower 
shapes in
the first sampling of the calorimeter in the region around the 
shower
direction defined by a $\Delta \eta \times \Delta \phi = 0.075\times0.025$ 
cluster. A shower generated by a $\pi^0$ is expected to be wider than a shower
generated by a single photon, and may have a second peak in its energy spatial
distribution. Several discriminating variables were 
defined~\cite{bib:PTDR,bib:wielers}:
\begin{itemize}
\item shower width on $n$ strips in strip units:
  $$\omega_{\mathrm{n}st}=\sqrt{\frac{\sum_i E_i 
    (i-i_{max})^2}{\sum_i E_i}},$$ where $i$ is an index 
  running over the
  selected number of strips, $i_{max}$ the strip corresponding 
  to the maximum energy deposit (in the present analysis, 
  shower widths over 3 and 21 strips were considered, called $w3st$ and $w21st$
  respectively) ;  
\item energy of the second maximum in the cluster, {\em e2max\/};
\item difference of the energy in the strip with the second maximum 
  and the energy deposit in the strip with the minimal value 
  between the first
  and second maximum, {\em edmax\/};
\item energy deposited outside of the shower core,
  $$eocore = \frac{E(\pm3) - E(\pm1)}{E(\pm1)},$$ where   $E(\pm n)$ 
  is the energy deposit 
  in $\pm n$ strips around (and including) the strip with 
  maximum energy deposit.
\end{itemize}
\par
\begin{figure} 
  \centering
  \includegraphics[scale=0.6]{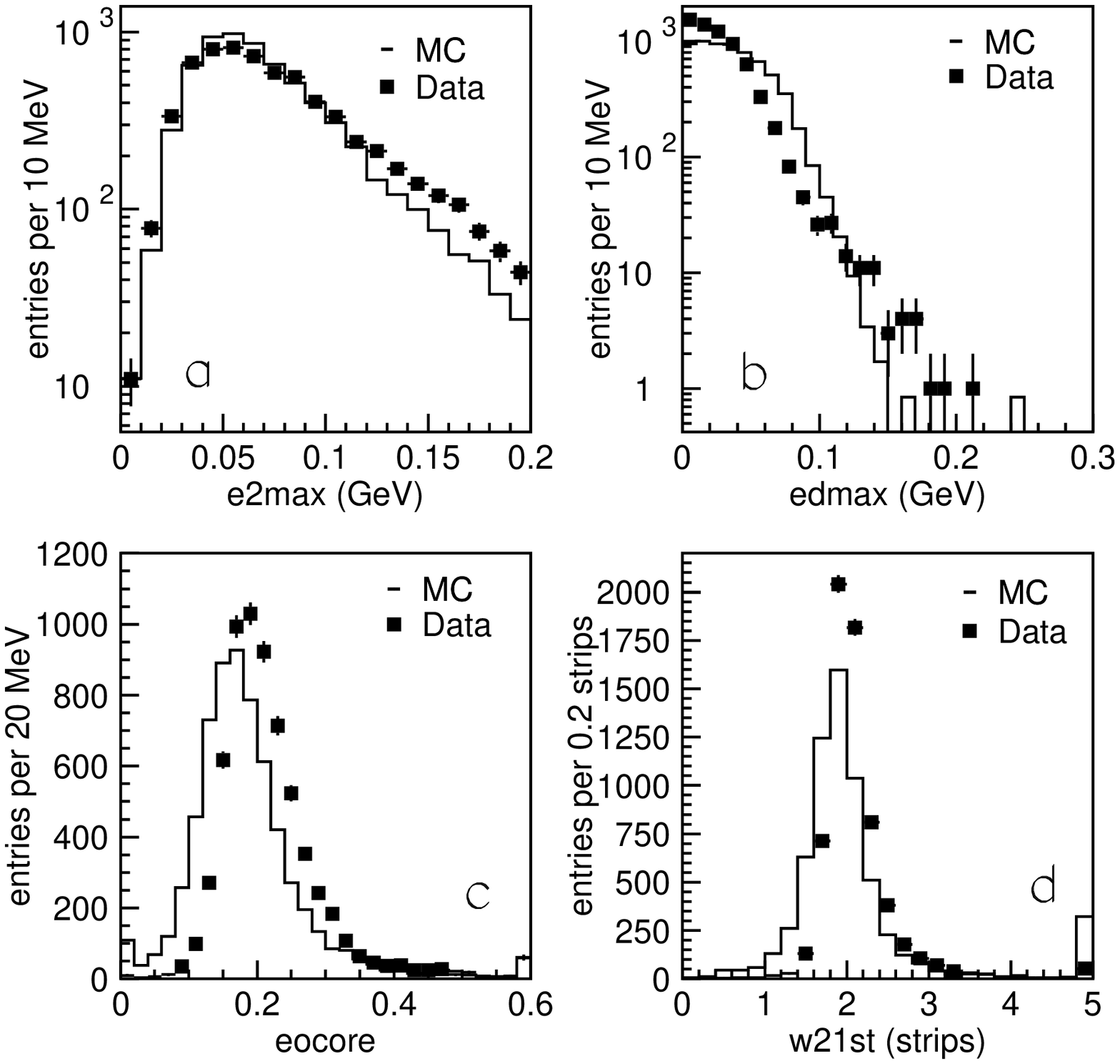}
  \includegraphics[scale=0.6]{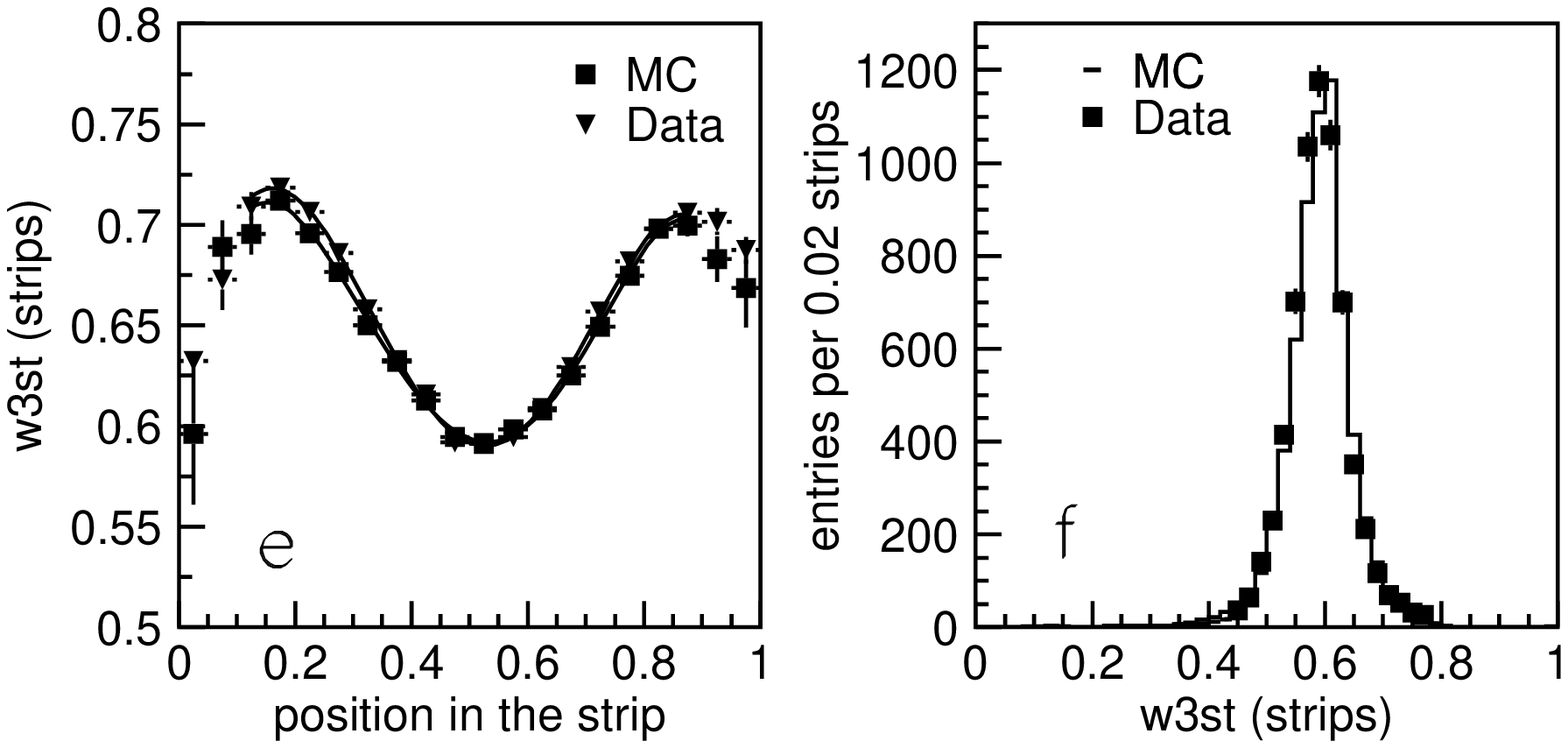}
  \caption{Discriminating variables for test-beam and simulated 
    photons: (a) energy of second maximum $e2max$, (b) energy difference between the minimum in the 
   valley and the energy of second maximum $edmax$, (c) energy deposited outside the shower core $eocore$, (d) 
   shower width computed over 21 strips $w21st$, (e)   shower width computed over 3 strips $w3st$ versus
   position in the cell and (f) $w3st$ after correction.} 
  \label{fig:phovar}
\end{figure}

The distributions of the discriminating variables, for samples 
of real and
simulated photons, are shown in Fig.~\ref{fig:phovar}. 
The choice of a cut
at 200~MeV for {\em e2max}, to reject multi-photon events, 
was justified by the
data/simulation discrepancy in the tails of both the {\em e2max\/} 
and {\em edmax\/} distributions; the $\pi^0$ rejection on the 
simulated sample was
not affected by this requirement.
As can be seen in Fig.~\ref{fig:phovar}-e, 
due to the finite granularity of the detector, the width
calculated on 3 strips depends on the impact 
position of the photon within the strip. This effect was corrected 
by fitting
the distribution of {\em w3st} versus the impact 
position of the photon within the strip, with a polynomial of 
degree 6; the
resulting {\em w3st} distribution is presented in 
Fig.~\ref{fig:phovar}-f.
\par
\begin{figure} 
  \centering
  \includegraphics[scale=0.6]{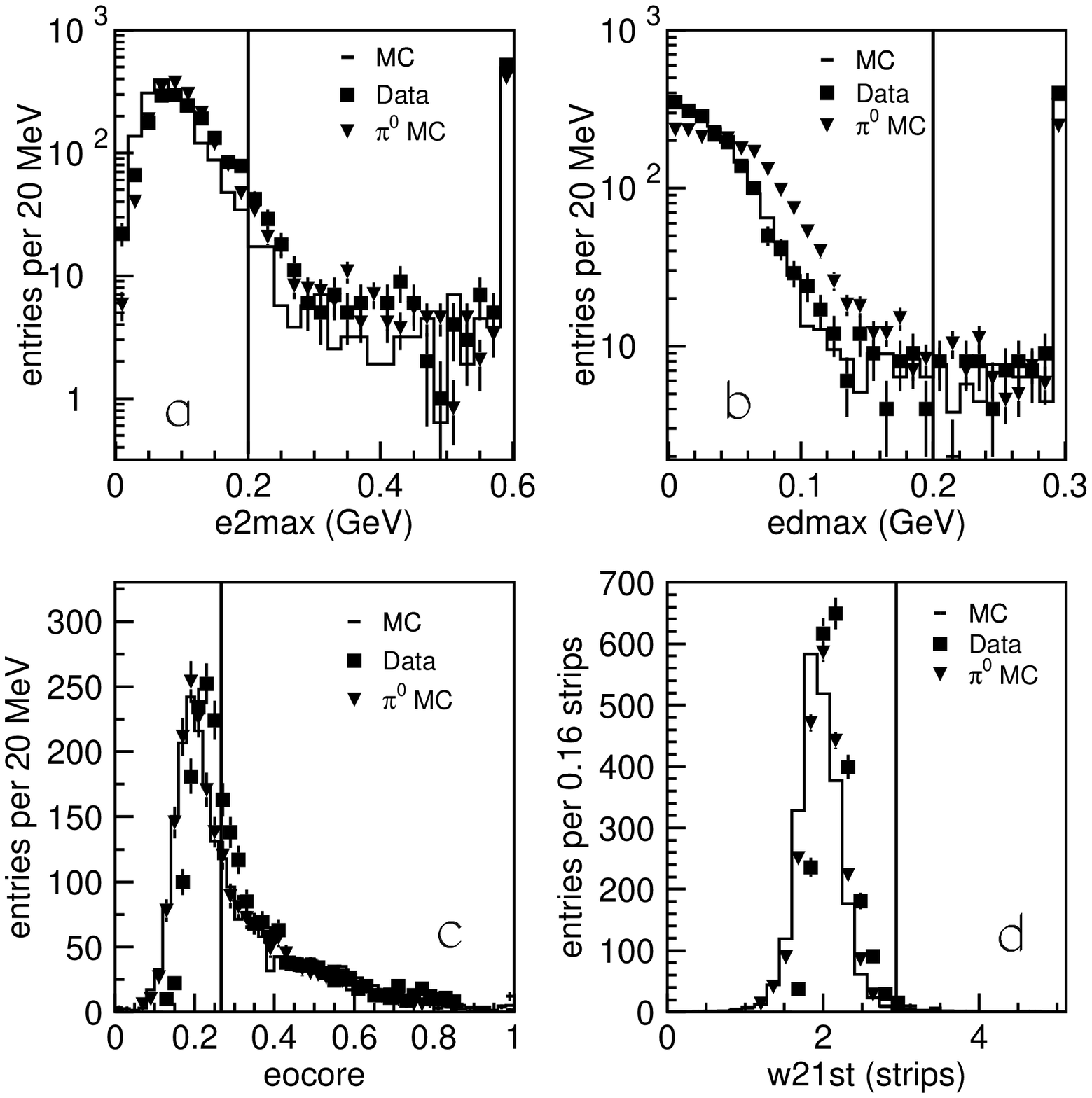}
  \includegraphics[scale=0.6]{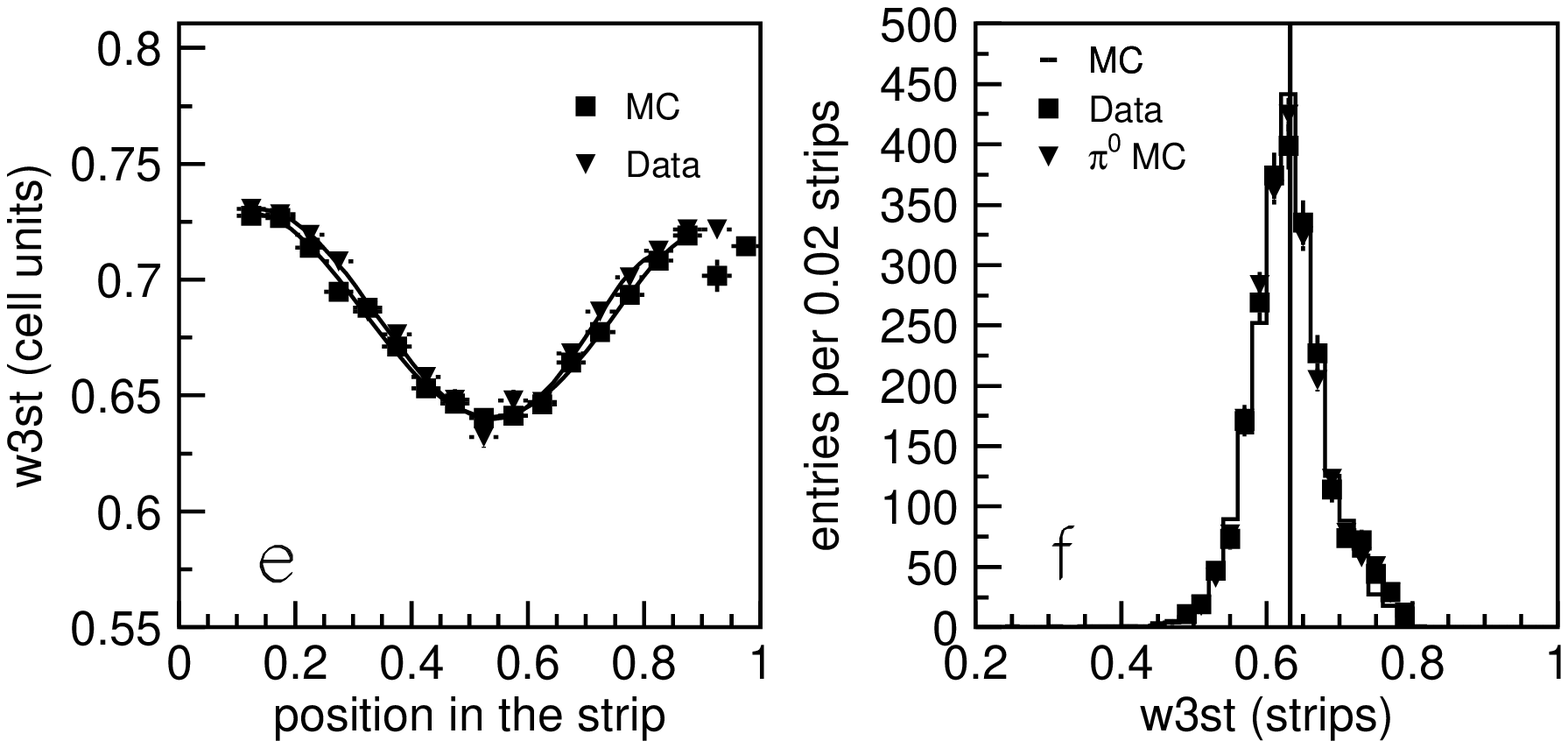}
  \caption{Discriminating variables for test-beam and simulated 
    pions: (a) energy of second maximum $e2max$, (b) energy difference between the minimum in the 
   valley and the energy of second maximum $edmax$, (c) energy deposited outside the shower core $eocore$, (d) 
   shower width computed over 21 strips $w21st$, (e)   shower width computed over 3 strips $w3st$ versus
   position in the cell and (f) $w3st$ after correction. The vertical lines define the cuts on each variable.}
  \label{fig:pivar}
\end{figure}

The same variables were calculated for the $\pi^0$-like
energy distributions, and are presented in Fig.~\ref{fig:pivar}. 
In the case of direct $\pi^0$ simulation (whose distributions in
Fig.~\ref{fig:pivar} 
are labelled as `$\pi^0$~MC'), the noise contribution 
was multiplied by $\sqrt{2}$.  
It can be seen that, due to the separation between the two photons, 
the average values for the distributions are larger for $\pi^0$-like
distributions than for photon-like ones. For the present analysis, 
a $\pi^0$
candidate was rejected when the values of the discriminating 
variables were
larger than the limits defined by the vertical lines in
Fig.~\ref{fig:pivar}. Sets of cuts were determined by 
independently varying the limit for each variable, while retaining 90\%
selection efficiency on the single photon subsample with $p_T=50$~GeV/c.
The set shown in Fig.~\ref{fig:pivar} was the one that yielded the 
best $\pi^0$ rejection.
%the final set of cuts was determined by 
%independently
%varying each value until the best $\pi^0$ rejection was
%achieved. 
\par
\begin{figure} 
  \centering
  \includegraphics[scale=0.5]{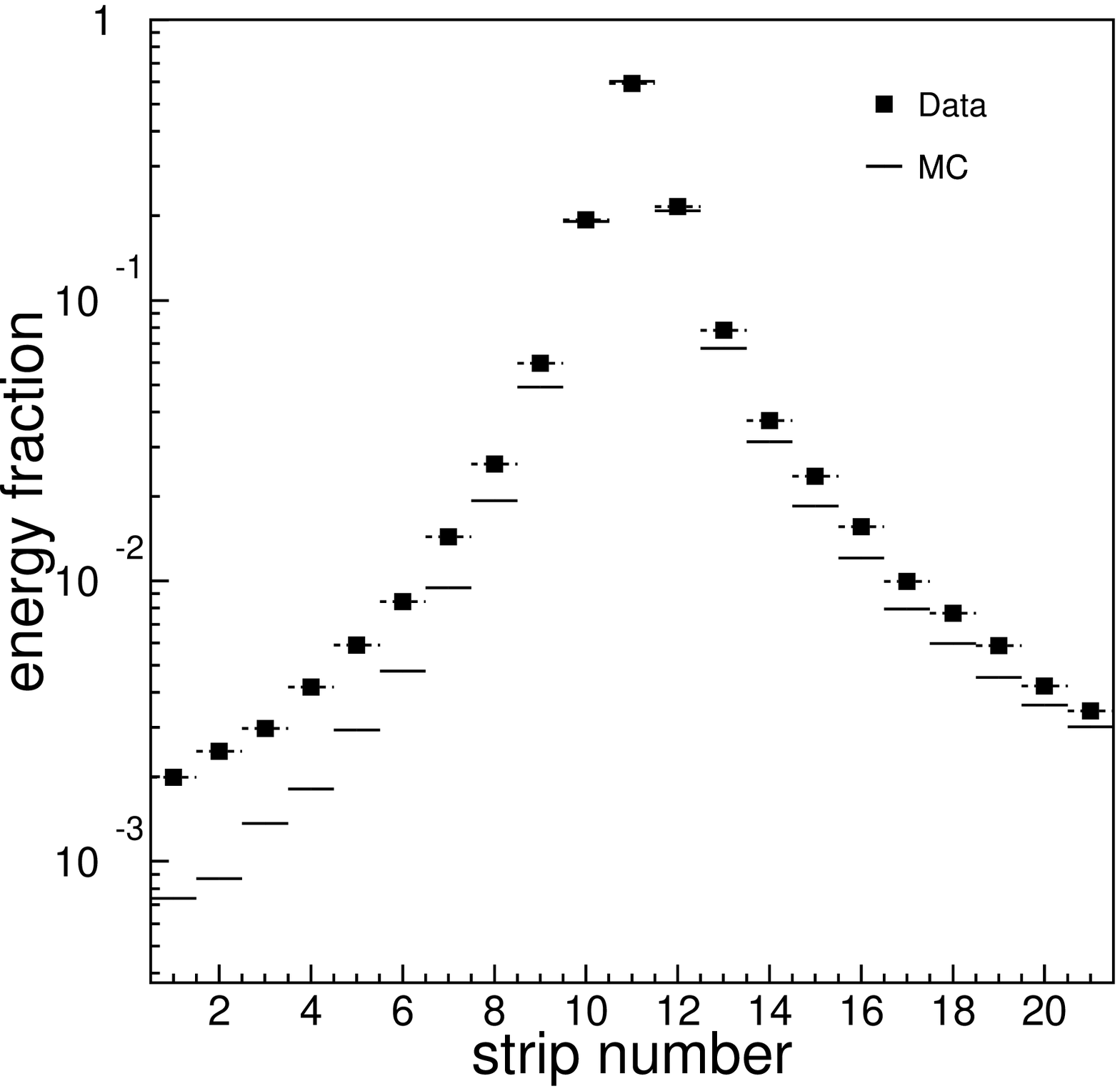}
  \caption{Lateral shower profile for data and simulation. 
    The energy
    fraction for each strip is plotted versus the strip number, 
    the shower
    maximum being at strip number 11. The energy fractions have 
    been normalised
    to the energy deposit in the 3 strips around the maximum.}
  \label{fig:lateral}
\end{figure}

The discrepancy between data and simulation, especially for the 
variables
{\em eocore} and {\em w21st}, can be explained by looking at the
data/simulation comparisons for the lateral shower profile.
In Fig.~\ref{fig:lateral} is shown that the agreement is good for 
the shower
core, i.e., the 3 strips around the maximum, but the discrepancy increases
with the distance from the maximum. The non-uniform distribution of dead
material located in the beam-line and not described in the Monte Carlo
might explain this discrepancy. 
\par
\begin{figure} 
  \centering
  \includegraphics[scale=0.45]{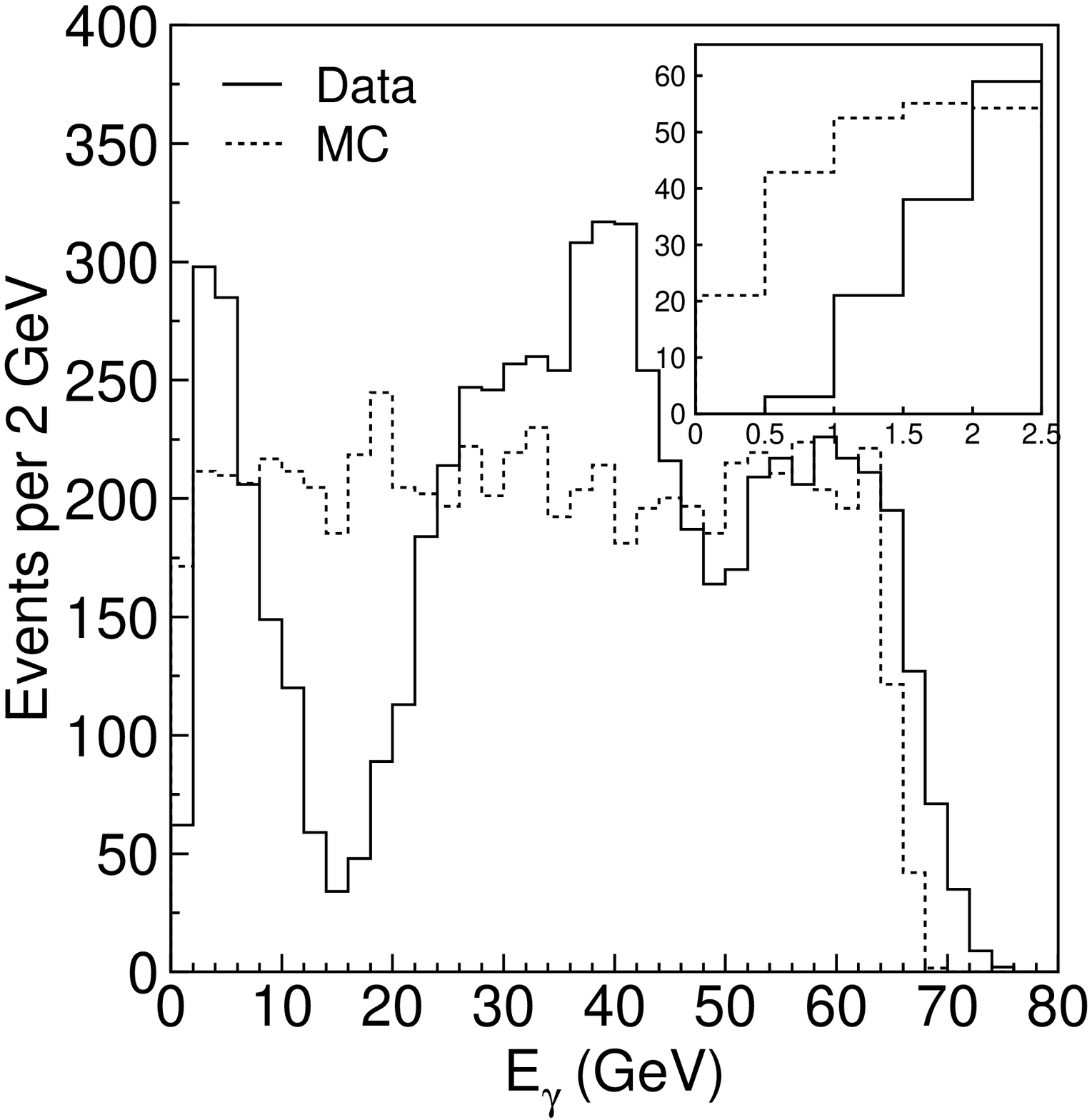}
  \caption{Photon energy spectrum for data and simulation. In the inset, 
the region between 0 and 2.5~GeV has been enlarged and a different binning
has been used.}
  \label{fig:energy}
\end{figure}

As only electron beam energy of 50 and
180~GeV were used to produce the photon beam, 
the resulting photon energy spectrum
was not uniform, but showed a lack of photons with energy around 
15~GeV,
as detailed in Fig.~\ref{fig:energy}. This affected the
reconstruction of pions shown in 
Fig.~\ref{fig:rejection}-a, where the variable
$\min(E_{\gamma_1},E_{\gamma_2})/E_{\pi^0}$, 
describing the
energies of the photons coming from the $\pi^0$ decay, is plotted for 
data and simulation. The lack of photons in the 
distribution for data is reflected by a large dip at 0.25: 
the $\pi^0$ rejection was therefore calculated in 6 bins of this variable, 
in the range 0 to 0.5.

\subsection{Results}

The photon efficiency was calculated for the photon sample at
a $p_T=$ 50~GeV/c. 
A bias in the calculation of the $\pi^0$ rejection was introduced
by the different starting points of the photon energy distributions 
for data and for the Monte Carlo shown in Fig.~\ref{fig:energy}. 
The number of photons is  
significant only for energies greater than 1~GeV in data, making it 
impossible to estimate the $\pi^0$ rejection for very asymmetric 
$\pi^0$ decays,
where one photon carries most of the pion energy. 
Assuming that the simulation correctly 
reproduces the photon spectrum at low energy, the
$\min(E_{\gamma_1},E_{\gamma_2})/E_{\pi^0}$ distribution for data 
can be extrapolated using the expected MC shape, and the rejection
recalculated taking this correction into account.
The difference between the rejection calculated in the 
second half of
the first bin of the $\min(E_{\gamma_1},E_{\gamma_2})/E_{\pi^0}$
distribution and the extrapolation to the full bin was -0.15.
%This can be corrected by assuming that the simulation correctly 
%reproduces the photon spectrum at low energy;
%the$\min(E_{\gamma_1},E_{\gamma_2})/E_{\pi^0}$ distribution for data 
%can thus be
%extrapolated using the expected MC shape, and the $\pi^0$ rejection 
%calculated for the first bin. 
\par
\begin{figure} 
  \centering
  \includegraphics[scale=0.65]{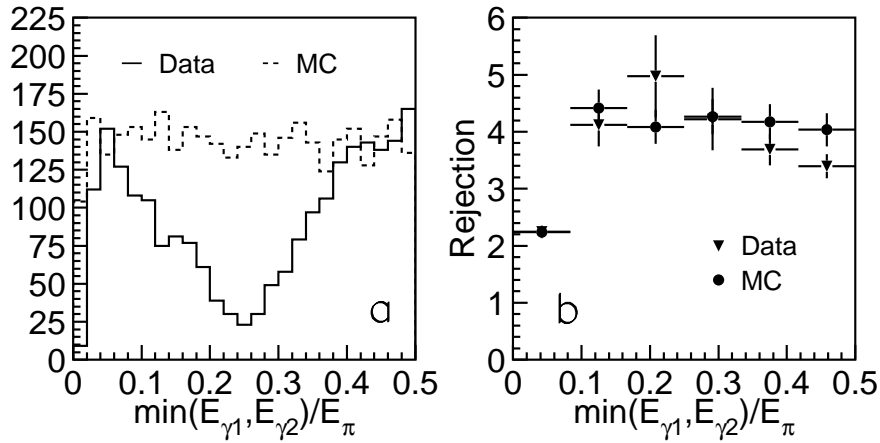}
  \caption{a) $\min(E_{\gamma_1},E_{\gamma_2})/E_{\pi^0}$
  distribution and b) $\pi^0$ rejection calculated in bins of
  $\min(E_{\gamma_1},E_{\gamma_2})/E_{\pi^0}$, for data and
  simulation.} 
  \label{fig:rejection}
\end{figure}

The distribution of the binned rejection factor is 
presented in
Fig.~\ref{fig:rejection}-b. 
The average value, obtained with a sample of 2300
reconstructed pions, was:
\begin{linenomath*}
\begin{equation}
R = \frac{n}{\sum_i \frac{1}{R_i}} = 3.54 \pm 0.12_{\mathrm{stat}}, 
\end{equation}
\end{linenomath*}
where $n$ is the number of bins and $R_i$ the rejection in each bin.
This value agrees with expectations in Ref~\cite{bib:PTDR}.

The same procedure was applied to simulated photons. 
They were paired up to
reproduce the kinematic simulation, and the requirements were
adapted to take into
account the differences between real and simulated distributions. 
In particular, the cuts were varied in such a way that the same 
fraction of
events was kept after each cut in data and simulation, resulting 
in an
overall photon selection efficiency of $(90\pm1)$\%. 
From the sample of
8000 simulated photons, about 3600 $\pi^0$ events were 
reconstructed. The
rejection factor , calculated using the same binning as for data, was
$R_{\mathrm{MC}}=3.66\pm0.10$. The good agreement of the calculated values of this
rejection factor for the data and Monte Carlo is 
shown in~Fig.~\ref{fig:rejection}-b 

\subsection{Systematic uncertainties}
\label{sec:syst}

The impact of the cross-talk  was studied by removing 
the 4.1\% cross-talk contribution described in 
section~\ref{sec:evsel}. The pion rejection degraded to $R = 3.28$, 
in good agreement with expectations~\cite{bib:tdr99}.
\par
The residual contamination of multi-photon events in the data sample
could introduce a bias in the determination of the $\pi^0$ rejection factor.
As the amount of matter before the
target was not well known, the contamination could be estimated 
by comparing
the standard sample, in which the converter on the photon beam 
was present,
with another sample where the converter was removed. 
Assuming that the
distribution of events with a certain number of photons 
is Poissonian, and
knowing the converter thickness and, therefore, the photon conversion
probability, it is possible to estimate the fraction of 
multi-photon events in
both samples by comparing the absolute number of counts. 
It was found that the
sample without converter contained 72\% of single photons, 
23\% of double
photons, 5\% of triple photons. For the sample with converter, 
these fractions were 84\%, 14\% and 2\% respectively. 
In principle, from these figures it should
be possible to extrapolate the value of the rejection to the 
case without
multi-photon contamination; in practice, due to the fact that 
no runs without
converter were taken at low energy, only a comparison for 
sub-samples could
be done. In fact, if one restricts the calculation of the 
rejection to the region  
$0.4 \leq \min(E_{\gamma_1},E_{\gamma_2})/E_{\pi^0} \leq 0.5$, 
both photons building the $\pi^0$ come from the sample
at high energy, for which the two setup conditions (with or without
converter) were present. The linear extrapolation to no 
background coming from
multiphotons induces a variation in the last bin of $+0.6\pm0.3$. 
\par
Finally, the pairing of the photons to build a $\pi^0$ was tested by
calculating the difference in rejection between the direct $\pi^0$
simulation and the $\pi^0$ Monte Carlo sample
obtained with the method described in section~\ref{sec:gammapi0rej}.
To allow a direct comparison between the samples, 
the noise level for the
direct $\pi^0$ simulation was multiplied by a factor $\sqrt{2}$.
The difference in the rejections was found to be +0.23.

%%%%%%

\section{e-$\pi$ separation}
\label{sec:epi}
Electromagnetic showers initiated by electrons are expected 
to be fully contained in the electromagnetic calorimeter. The hadronic 
showers start at a larger depth of the module and there is often a 
substantial fraction of the total energy shower leaking into 
the hadronic calorimeter. However, a fraction
of hadron-initiated showers may be fully contained
creating a potential for particle misidentification. 
Therefore it is necessary to use the longitudinal and lateral 
segmentation of the
electromagnetic calorimeter to minimise pion misidentification as electrons while
maintaining high electron identification efficiency.

\subsection{Clustering and shower profiles}
Data were analysed using the standard clustering procedure described in
section~\ref{sec:posres}.
The shape of the longitudinal shower profile was
contained in the
information of the energy $E_i$ deposited in each layer of 
the calorimeter.
Additional
information was contained in the lateral shower profile, 
characterised by the
number of ``hit'' cells in each layer,  NH($i$), i.e. the number of
cells that 
contain energy above the noise level. 
For each cell the noise level was
obtained using the random trigger. The noise distribution 
was fitted with a
Gaussian function and the noise level was given by the standard 
deviation of the
fit. It was found to correspond to 44, 12, 32, 21~MeV respectively for 
the four samplings of the calorimeter in this rapidity region. 
The minimum energy required for a cell to register a hit 
corresponded to a $5\sigma$ cut above the noise level. 
\par
\begin{figure}
  \centering
  \includegraphics[scale=0.5]{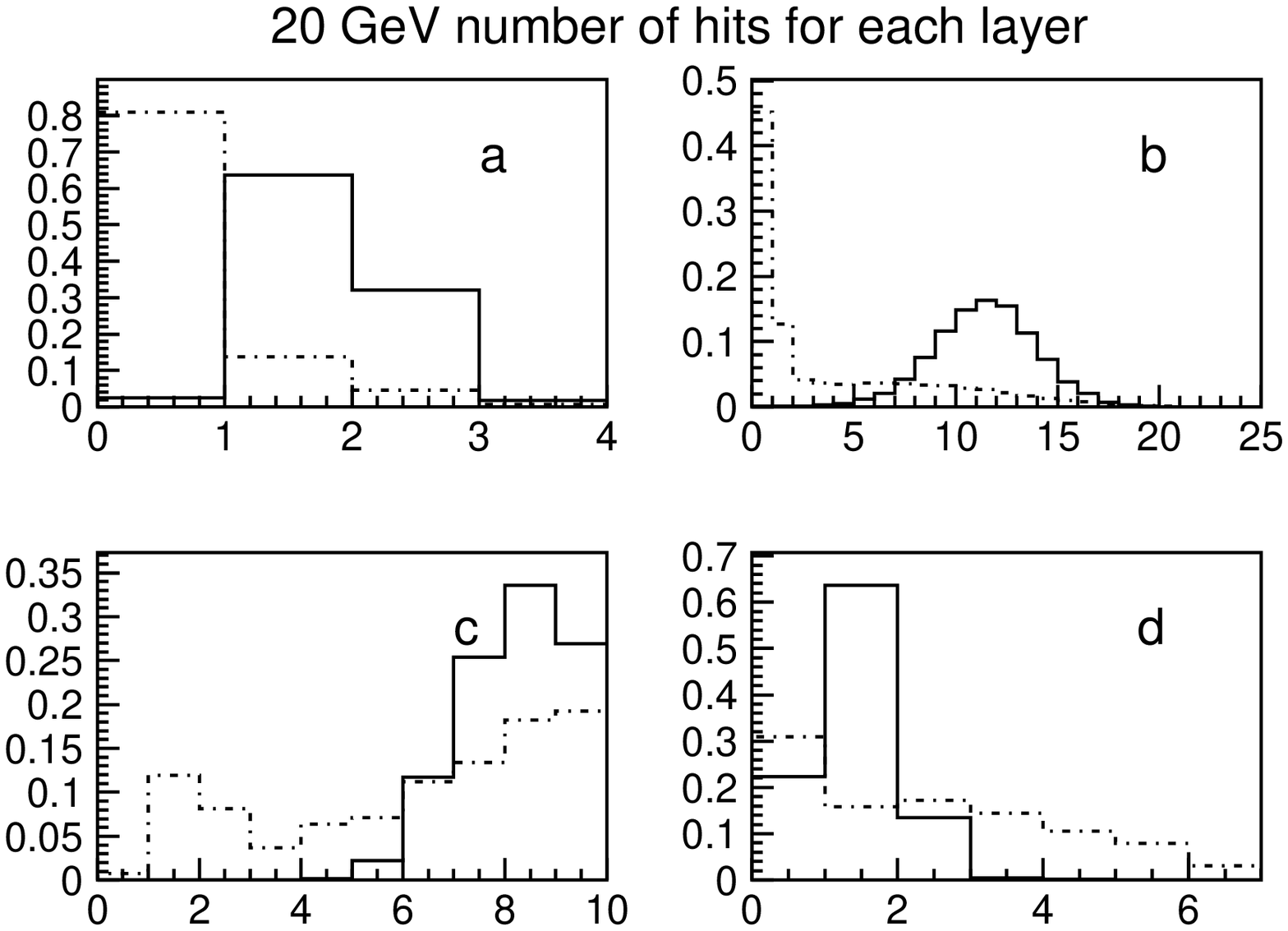}

%  \vspace{0.4cm}

%  \includegraphics[scale=0.5]{title-40-nhits.eps}
  \includegraphics[scale=0.5]{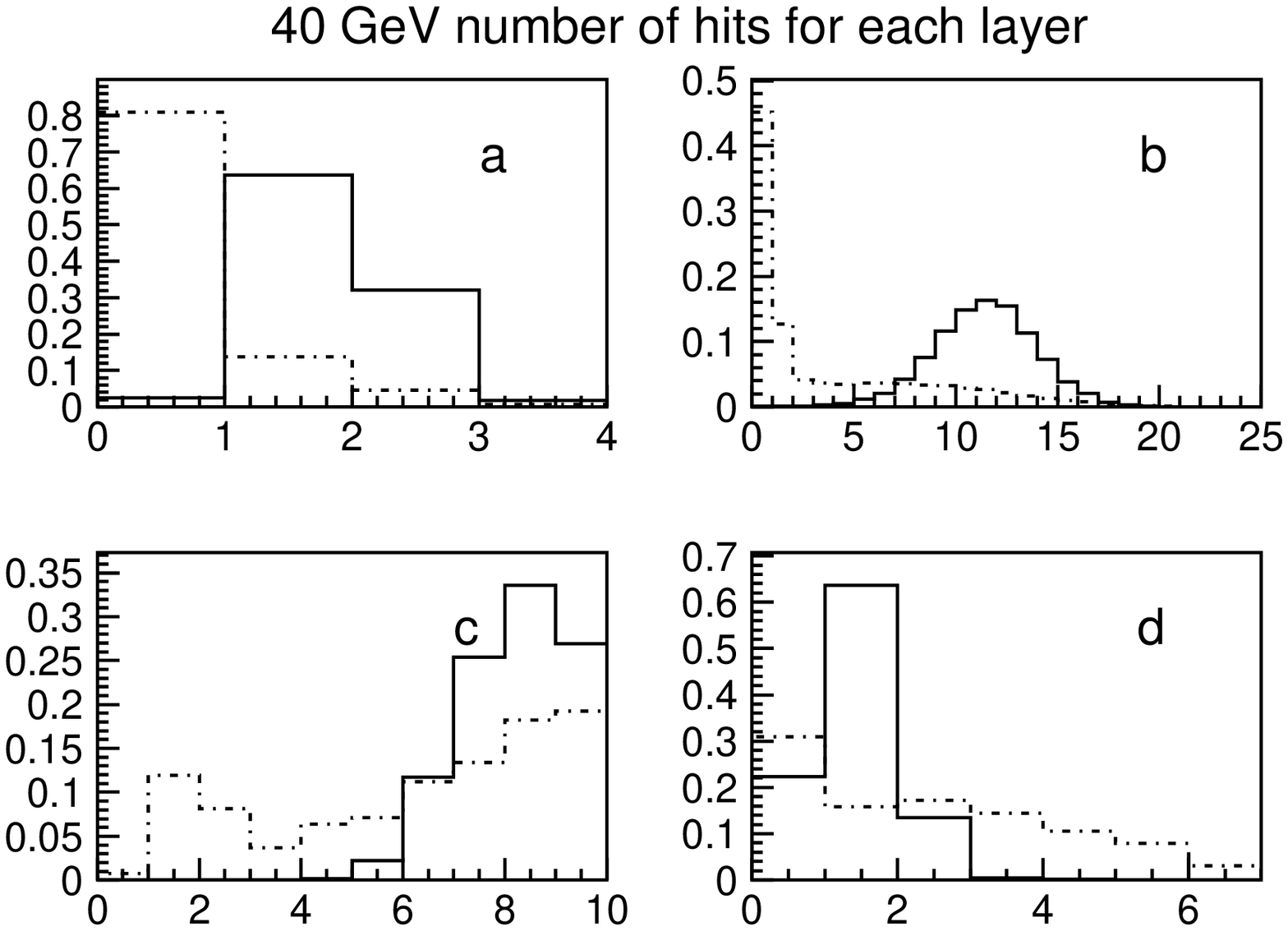}
  \caption{Number of hit cells for 20 and 40~GeV/c 
    electrons and pions for a)
    presampler, b) front sampling, c) middle sampling 
    and d) back sampling of
    the EM calorimeter. All plots are normalised to 1. 
    Solid line stands for
    electrons and dashed line stands for pions.}
  \label{fig:hitcells}
\end{figure}

In this analysis, the number of hit cells in a given layer was 
calculated only within the selected cluster. 
The number of hit cells is quite different
for electrons and
pions, 
for the 20 and 40~GeV/c beams, as shown in Fig.~\ref{fig:hitcells}, 
and may provide additional discrimination power.
\par
\begin{figure}
  \centering
  \includegraphics[scale=0.6]{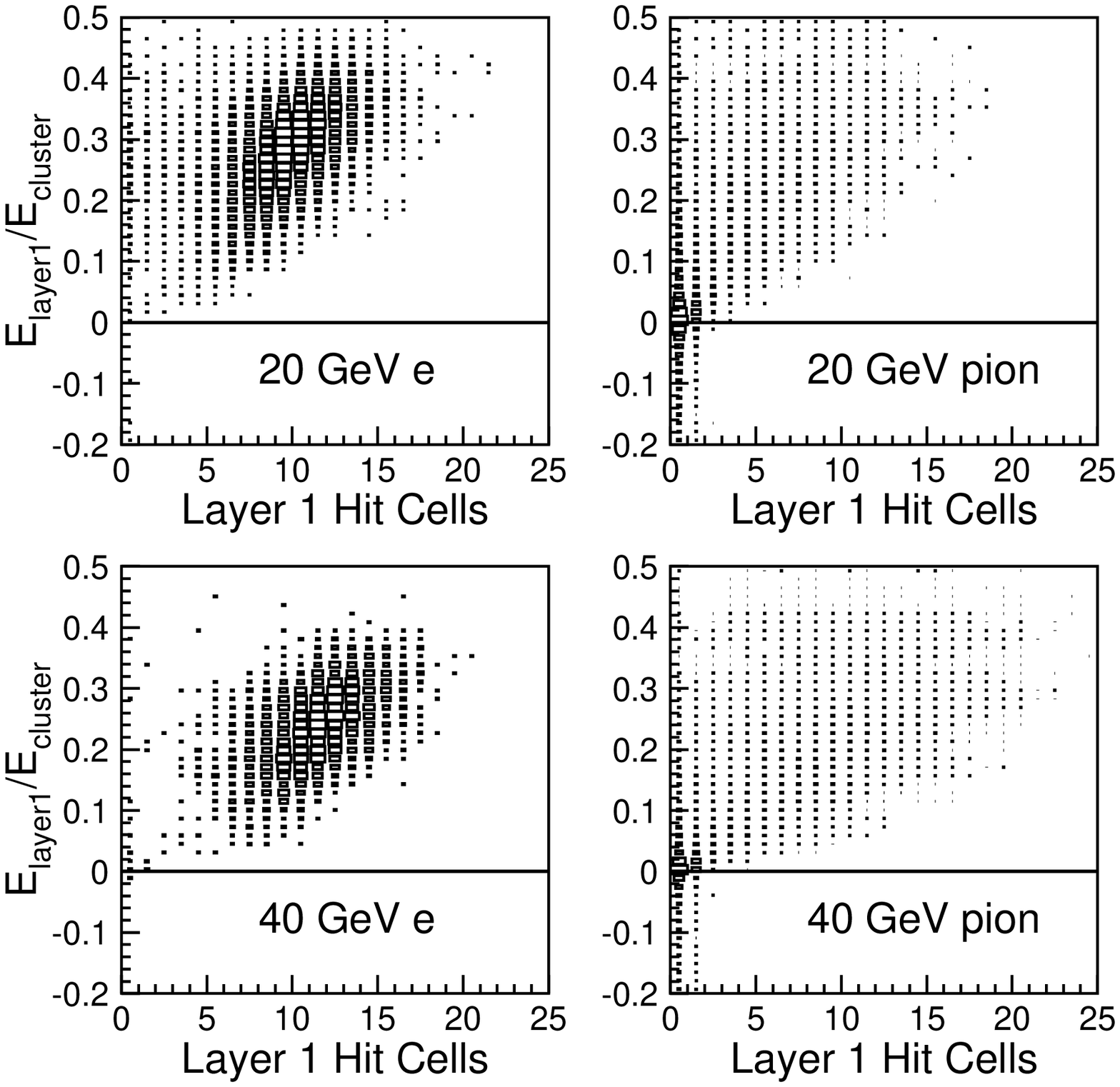}
  \caption{Number of hit cells in the front sampling (Layer 1) vs
  $E_{\mathrm{Layer 1}}/E_{\mathrm{Cluster}}$ for 20 and 40~GeV/c 
    electron and
    pion samples.}
  \label{fig:correl}
\end{figure}

Instead of total energy deposits in the calorimeter, 
ratios of the energy in each individual layer 
to the total energy deposited in the cluster, $E_i/E_{\mathrm{Cluster}}$, 
were used.
In general, the number of hit cells and the energy fraction 
in a given layer
are correlated. 
This correlation shown in
Fig.~\ref{fig:correl} is weak. Thus the number of hit cells provides additional 
information 
%that can give the same efficiency for electrons' selection 
%as with 
to a longitudinal profile analysis. The additional information allows for
significantly better pion rejection.
Another variable used in the analysis is the fraction of 
energy deposited in
the shower core. To obtain it, the central cell of
the middle sampling cluster was projected onto the other layers, 
and the
total energy $E_{11}$ corresponding to such calorimeter 
tower calculated. The
fraction of energy deposited in the shower core was then defined as
$E_{11}/E_{\mathrm{Cluster}}$.

\subsection{Sample purity}
To select the electron sample, the 
ADC counts from the \v{C}erenkov counter were required to be 
greater than 400,
above the pedestal, 
Pion run samples were
normalised to the electron run samples with ADC counts less than 400
according to their total area. 
The tail of the ADC counts distribution of the
pion sample above 400 gave the maximum contamination in the electron
sample. The purity of the electron sample was defined as 
the number of
electron sample counts with pion contamination subtracted, 
divided by the
number of electron sample counts, and was found to be 
better than 99.9\%.
\par
The pion sample was selected by requiring the \v{C}erenkov 
ADC counts to be less 
than 400 for the pion run data, and its purity was 
calculated to be better
than 99\% for all momenta.

\subsection{Analysis strategy}
The variables used in the analysis are the four energy fractions
$E_i/E_{\mathrm{Cluster}}$, the fraction of energy deposited in 
the shower
core $E_{11}/E_{\mathrm{Cluster}}$, and the number of hit cells 
in each layer NH($i$).
Assuming that the track momentum $p$ is known, one may 
use it as
additional variable, and study $e-\pi$ separation with samples 
of the same momentum.
\par
Two analysis methods~\cite{bib:epi} made use of the variables defined so far. 
The first
method was based on simple cuts to separate electrons of pion signals
while keeping the electron identification efficiency at about 90\%.
The second method relied on a Neural Network (NN) approach
using the same variables. In particular, a
multi-layer perceptron (MLP) with one hidden layer was used to 
build the
NN. Monte Carlo data were divided into three equal subsamples of 
a specific
momentum and particle type, serving as NN training sample, 
verification
sample and a testing sample. 
The number of neurons varied with the number of
input variables, following the rule of thumb that the 
number of hidden neurons
should be about twice the number of input variables. 
The  neural network
results were stable when changing the number of hidden neurons. 
The error of
the output of the NN was checked to avoid over-training.

\subsection{Results}
The results are summarised in Tables~\ref{tab:20gev} and~\ref{tab:40gev} for 
analyses  based on cuts and Neural Network approach, 
applied to 20 and 40~GeV electron and pion samples. The 
pion fake rates for fixed electron efficiency of 90 \% are compared 
for the two methods. For the NN method, a single cut applied on the 
NN output provides the result. For the cut method, the thresholds were varied on
all variables in order to minimise the pion fake rates while preserving the electron
identification efficiency.

\begin{table}[ht!]
  \centering
  \begin{tabular}{ccc}\hline\hline
    20~GeV with 90\% electron efficiency & Fake rate (CUT) 
                                            & Fake rate (NN)
    \\ \hline
    Energy ratio & $(6.88\pm0.27)$\% & $(4.07\pm0.29)$\% \\
    Energy ratio + hit cells & $(3.18\pm0.18)$\% 
                                     & $(2.42\pm0.22)$\% \\
    Ratio + hit cells + $E_{\mathrm{Cluster}}/p$ 
                                & $(0.50\pm0.07)$\% &
      $(0.34\pm0.08)$\% \\
   \hline\hline
  \end{tabular}
  \caption{Results for 20~GeV samples using simple cuts (CUT) 
    and Neural Network (NN) methods.} 
  \label{tab:20gev}
\end{table}

\begin{table}[ht!]
  \centering
  \begin{tabular}{ccc}\hline\hline
    40~GeV with 90\% electron efficiency & Fake rate (CUT) & Fake rate (NN)
    \\ \hline
    Energy ratio & $(6.35\pm0.15)$\% & $(3.52\pm0.14)$\% \\
    Energy ratio + hit cells & $(4.26\pm0.12)$\% 
                              & $(1.88\pm0.10)$\% \\
    Ratio + hit cells + $E_{\mathrm{Cluster}}/p$ 
                              & $(0.38\pm0.04)$\% &
                                 $(0.07\pm0.02)$\% \\
   \hline\hline
  \end{tabular}
  \caption{Results for 40~GeV samples using simple cuts (CUT) 
    and Neural Network (NN) methods.} 
  \label{tab:40gev}
\end{table}

\section{Conclusions}
Test-beam data collected in the years 2000-2002 with 
pre-series and series
modules of the ATLAS EM barrel and end-cap 
calorimeter have shown that the
design goals in terms of position and polar angle
resolution, $\pi^0$ rejection and $e-\pi$ separation can be reached.
The position resolution along $\eta$ was measured to be about
$1.5\times10^{-4}$ and $3.3\times10^{-4}$ (in units of pseudorapidity) 
for front and middle compartment
respectively, allowing to achieve a polar angle 
resolution in the range
$50-60~(\mathrm{mrad})/\sqrt{E~(\mathrm{GeV})}$. 
The $\pi^0$ rejection, in test-beam conditions, 
was calculated to be $3.54
\pm 0.12_{\mathrm{stat}}$ for 90\% photon selection efficiency at
$p_T=50$~GeV/c, with room for improvement when the systematic 
errors from irreducible multi-photon background ($+0.6\pm0.3$) and 
$\pi^0$ building strategy (+0.23)
are taken into account.
Finally, the study of $e-\pi$ separation performed at 
20 and 40~GeV yielded a
fake rate of $(0.50\pm0.07)$\% and $(0.38\pm0.04)$\% 
with a simple cuts
method, $(0.34\pm0.08)$\% and $(0.07\pm0.02)$\% with a Neural
Network method, 
while maintaining in both cases 90\% electron identification
efficiency. Significant improvement in $e-\pi$ separation can be achieved 
by using the lateral shape of the shower described by the number of hit cells.

\section*{Acknowledgements}
We are indebted to our technicians for their 
contribution to the construction
and running of the barrel and end-cap modules and the electronics. 
We would like to thank the accelerator division for the good working
conditions in the H6 and H8 beamlines. 
Those of us from non-member states
wish to thank CERN for its hospitality. 

% \appendix

% \section{}
% \label{}

\end{document}